\documentclass{article} 
\usepackage{amsmath,amsopn,amscd,amssymb,xypic,rotating,epic}
\xyoption{all}

\newcommand{\double}       {}
\newcommand{\single}       {}

\def\shortcite{\cite}

\newcommand{\obs}[1] {\begin{center}\framebox{\parbox{0.75\columnwidth}{\textbf{#1}}}\end{center}}
\newcommand{\argmin}{\arg\!\min}
\newcommand{\argmax}{\arg\!\max}

\newcommand{\VInsert}[2]   {\centerline{\immediate\pdfximage height #2 {#1}\pdfrefximage\pdflastximage}}
\newcommand{\HInsert}[2]   {\centerline{\immediate\pdfximage width #2  {#1}\pdfrefximage\pdflastximage}}

\newcommand{\cmd}[1]      {\underline{{#1}}}
\newcommand{\cb}          {\begin{tabbing}MMMMM\=MM\=MM\=MM\=MM\=MM\=MM\=MM\=MM\=MM\= \kill}
\newcommand{\ce}          {\end{tabbing}}

\newcommand{\tset}[1]      {\{{#1}\}}                     
\newcommand{\tbag}[1]      {\{\hspace{-0.25em}|{#1}\}\hspace{-0.5em}|}         
\newcommand{\tlist}[1]     {[{#1}]}                       
\newcommand{\ttree}[1]     {{\frak t}({#1})}              

\newcommand{\separate}     {\vspace{0.3cm}\begin{center}*~~~~~~~~~~*~~~~~~~~~~*\end{center}\vspace{0.3cm}}
\newcommand{\dqt}[1]        {"{#1}"}
\newcommand{\bp}            {\rule{5em}{0pt}}
\def\emph{\textsl}
\def\em{\sl}
\def\textbf{\pmb}

\def\som{M}
\newcommand{\nsym}[1]   {\textbf{\texttt{{#1}}}}

\newcounter{myremark}
\newcounter{myexample}
\def\example{
\bigskip

\refstepcounter{myexample}%
\noindent \textbf{Example \Roman{myexample}:}\\
}

\def\remark{
\refstepcounter{myremark}%
\noindent \emph{Remark \arabic{myremark}:}
}

\renewcommand{\topfraction}{.99}
\renewcommand{\textfraction}{.01}

\def\capstyle{\scriptsize}

\begin{document}

\title{Novelty and Coverage in context-based information filtering}
\author{Alexandra Dumitrescu and Simone Santini}
\date{Escuela Polit\'ecnica Superior\\Universidad Aut\'onoma de Madrid}

\maketitle

\double

\begin{abstract}
  We present a collection of algorithms to filter a stream of
  documents in such a way that the filtered documents will cover as
  well as possible the interest of a person, keeping in mind that, at
  any given time, the offered documents should not only be relevant,
  but should also be diversified, in the sense not only of avoiding
  nearly identical documents, but also of covering as well as possible
  all the interests of the person. We use a modification of the WEBSOM
  algorithm, with limited architectural adaptation, to create a user
  model (which we call the \emph{user context} or simply the
  \emph{context}) based on a network of units laid out in the word
  space and trained using a collection of documents representative of
  the context.

  We introduce the concepts of \emph{novelty} and
  \emph{coverage}. Novelty is related to, but not identical to, the
  homonymous information retrieval concept: a document is novel it it
  belongs to a semantic area of interest to a person for which no
  documents have been seen in the recent past. A group of documents
  has \emph{coverage} to the extent to which it is a good
  representation of all the interests of a person.

  In order to increase coverage, we introduce an \emph{interest} (or
  \emph{urgency}) factor for each unit of the user model, modulated by
  the scores of the incoming documents: the interest of a unit is
  decreased drastically when a document arrives that belongs to its
  semantic area and slowly recovers its initial value if no documents
  from that semantic area are displayed.

  Our tests show that these algorithms can effectively increase the
  coverage of the documents that are shown to the user without
  overly affecting precision.
\end{abstract}

\section{Introduction}
The ready availability of data made possible by modern digital
communication systems and the sheer amount of data that these system
make accessible have created a curious inversion in the relation
between people and the information they seek. For centuries, the basic
information problem was to find enough of it. The press, the public
library, the newspapers are devices designed to solve or alleviate
this problem, reaching an uneasy and unstable balance between the
amount of data we were exposed to and the useful information that we
were capable of getting out of them. Digital storage and digital
communication have thrown this balance awry: the problem that we (and,
very likely, the future generations) face is no longer to get to the
data, but to avoid being overwhelmed by them; no longer searching for
the precious stuff, but keeping most of them out of our (digital)
door.

For centuries, the relation between data, information, and knowledge
was considered---and with good reason---a direct one: one could get better
knowledge from more information, and one could get more information by
having access to more data. Techniques like
statistics or data visualization \cite{tufte:86}, which distilled, so to
speak, information from data, do not impinge on this basic tenet: data
are a scarce and valuable resource, and one should look at and analyze
all the data that she can put her hands on. 
Consequently, for centuries, laying one's hand on enough relevant
documents has been a central problem for whoever pursued an
intellectual interest. Around the end of the X century Gerbert (later
Pope Sylvester II) wrote

\begin{quote}
  I am working on the creation of my library. For a long time in Rome
  and in all Italy, in Germany and in Belgium I spent great amounts of
  money to pay copyists and buy books, helped by the solicitude and
  benevolence of my friends. Allow me then to pray you to render me
  the same service. According to what you say, I will send the
  copyist, the parchment, and the necessary money, and I will be
  grateful for your help \cite{barthelemy:68}
\end{quote}

Casting this situation in the language of information retrieval, we
could say that the emphasis of data acquisition was preponderantly on
\emph{recall}: in a situation of scarcity of data, the paramount
preoccupation is to avoid missing any potentially useful document, and
the risk of having to analyze some irrelevant ones is a small price to
pay for this guarantee.

Things have changed. The amount of data available on the internet is
such that not only is recall no longer a primary concern: it has
become virtually unmeasurable, if not meaningless. Any set of
retrieved documents small enough to provide useful information (viz.,
to be manageable by one person) will contain only an insignificant
fraction of what is available on-line, and its recall will therefore
be practically zero. Conversely, any set of documents with a
significant recall will, due to the sheer amount of data available, be
too large to provide any useful information. At the same time, 
despite the
amount of data available, we don't seem to be better informed than we
used to: drowning in a plethora of trivia, it is easy to overlook the
things that we truly would have cared about.

So, the emphasis has shifted to trying to achieve high
\emph{precision}, that is, to retrieve only (or mainly: nobody's
perfect) relevant documents.

Precision, however, is only part of the story due to another
frustrating characteristic of ultra-large collections such as the
internet: the presence of many documents that say pretty much the same
thing. Much like in Borges's \emph{Babel Library}, to which the
internet is the best approximation so far, for any given document on
the internet there are likely to be hundreds more that, without being
quite the same, say more or less the same things.

Data are not information, and the relation between them is non-linear
and highly contextual. A document may be highly informative if it is
the first document one receives on a given topic, but its information
content (which, unlike its data content, is subjective and relative to
the reader) is virtually nil if we see it after seeing another
document that contains more or less the same data. In other words: the
information content of a document is not simply a function of its data
content, but depends on the data content of the other documents that a
person is examining. Observations such as these have caused a shift of
emphasis, in Information Retrieval, from systems based purely on
relevance to systems that take into account further dimensions such as
\emph{novelty} and \emph{diversity}. The classical Robertsonian model
of Information Retrieval \cite{robertson:77,robertson:76} assumed that
the relevance of a document for a query is a function of the query and
the contents of the document. Beginning around the turn of the
century, this assumption has been criticized based on considerations
similar to those we just made. A result set of highly relevant
documents (in the Robertsonian set) that contains very similar
document, each one relating more or less the same information, would
not be very informative \cite{clarke:08,clarke:09}. Novelty and
diversity are measures that attempt to avoid this kind of situation. A
document is \emph{novel} with respect to a set of documents, if it
contains information not present in any other document of the set; a
set of results is \emph{diverse} if it covers all different aspects of
a query.

In all these cases, novelty and diversity are measures relative to
\emph{one} result set. This is because in the typical model of
Information Retrieval, each query is an independent interaction, so
the current result set is the only possible context with respect to
which the novelty of a document can be measured.

In this paper, we consider a different scenario: that of filtering
\cite{oard:97} continuous \emph{streams} of documents (more specifically, news items)
so that incoming documents of potential interest are shown to the
user, while documents of scarce interest are filtered out. The
considerations that led researchers to develop the concepts of novelty
and diversity are valid here as well: if an important event takes
place, various sources will talk about it, often repeating more or
less verbatim the same press release. Presenting over and over again
similar documents on the same topic will soon result in data overflow
and poor information. This common root notwithstanding, two
characteristics of our scenario suggests that we should take a
somewhat different approach to the diversification of results:

\begin{description}
\item[i)] In Information Retrieval, novelty and diversity are defined
  for finite result sets, but in our case we are in the presence of a
  continuous flow of results that change as new items arrive. 
\item[ii)] A filtering system doesn't deal with a specific information
  need clearly defined by a query, the way an Information Retrieval
  system does. Consequently, when we consider novelty and diversity,
  we must refer these terms not to a result set but to the general
  interests of a person and, at the same time, we can't ignore the
  temporality of the results: a document is novel if it covers a topic
  of interest that hasn't \emph{recently} been covered by other
  documents.
\end{description}

The solution of our filtering problem requires two elements: on the
one hand, we must build a suitable model of a person's general
interests; on the other hand, we must be able to use this model to
extract, from the stream of documents, those that cover topics of
interest for the person that haven't been covered for some time. The
time component is essential here: a document may not be considered
novel if the topic it covers has been covered \emph{recently} by
another documents. However, the user interest that has been satisfied
by a previous document will arise again if the topic hasn't been
covered for some time, so the same document may be considered novel if
it is presented again some time later, when its topic hasn't been
covered in the recent past.

We present and evaluate the algorithmic foundations of a news
filtering system based on two components.

\begin{description}
\item[i)] A dynamic model of a person interest. We use the documents
  with which a person interacts to train a self-organizing network
  that will act as a \emph{latent semantic manifold}. Incoming
  documents are represented using a vector space representation, and
  their distance from the latent semantic manifold is used as a
  measure of relevance for the user interest.
\item[ii)] A mechanism for taking into account the novelty of
  items. Whenever an item is shown that is close to a region of the
  latent semantic manifold, the \emph{urgency}, or \emph{interest} of
  that region is reduced during a certain amount of time so that
  further documents too similar to the one already shown will have
  their relevance reduced.
\end{description}

We present our user model in Section 2, in which we also introduce the
basic filtering algorithm, that is, the algorithm that selects
documents of potential interest to the user without taking into
account novelty and diversity.

Our new scenario---continuous flow of items and context represented by
the user model---forces us to analyze afresh the concepts of novelty
and diversity, an analysis that we carry out in Section 3.  As a
result, we maintain the notion of novelty, albeit in a somewhat
different form that the usual one, but we replace diversity (a
property of a result set) with \emph{coverage}: a measure of how much
do the results cover all the interests of a person. In Section 4, we
extend the algorithm to take into account novelty and to increase
coverage.  In Section 5, we present our testing methodology and in
Section 6 we present our results.  Related work is discussed in
Section 7, and some conclusions are drawn in Section 8.

Before we conclude this introduction, we should like to make two
observations. Firstly, our user model is based on the analysis of
documents of interest to the user, but we are agnostic with respect to
the origin and nature of these documents: files in the computer,
emails, texts, queries to data bases, search history, etc. The
selection of suitable sources is a system design issue that we do not
consider here.

Secondly, we filter by analyzing only the text of the news, that is we
ignore the possible presence of meta-data in the stream. Meta-data are
a valuable source of information, and they obviously play an important
r\^ole in the creation of an information system. They are, however,
often determined \emph{a priori}, not relative to the interests of the
user, and not discriminating enough, so that they often need to be
supplemented with algorithms that work on the actual contents of the
documents. Such algorithms form the subject of this paper.

\section{The User Model}
\label{sec:somActiveApproach}
Our user model, which we introduce in this section, is an adaptation
of our previous work \cite{santini:09a} which in turn is based on
self-organizing maps (SOM, \cite{kohonen:90}) and their application to
information retrieval (WEBSOM, \cite{kaski:97}).

The base of our model is the standard vector space of
information retrieval \cite{salton:88}, in which each axis represents a word
(more precisely: a \emph{stem}--see the following). A point in this
space is a vector $p=(p_1,\ldots,p_W)$ (where $W$ is the number of
words) and can be loosely considered to represent a concept, $p_i$
being the degree to which the word $i$ is part of the concept $p$. For
technical reasons, we shall work with normalized vectors
($\sum_ip_i^2=1$), so that we shall not work on the whole word space
but on the unit sphere in this space.

Our context representation is a modification of the standard SOM
algorithm as it is normally used in information retrieval. Before
introducing our modified version, we consolidate the terminology by
introducing briefly the standard algorithm.

\subsection{The standard SOM algorithm}
\label{standardSOM}
A Self-Organizing Map constituting a context, thus as it was used, for
example, in \cite{kohonen:90} consists of a two dimensional grid of
\emph{units}%
\footnote{These units receive different names in the literature: given
  their grid arrangements they are sometimes called \emph{nodes}, and
  in the neural network literature they are in general called
  \emph{neurons}. Here we shall use the neutral term \emph{units}.}%
, arranged in a rectangular $N\times{N}$ grid, where each unit,
$u^{i,j}$, is a point in the word space:
\begin{equation}
  u^{i,j} = [u_1^{i,j}, u_2^{i,j}, \ldots, u_W^{i,j}]'
\end{equation}
The units are related to each other as points in the word space (using
some distance $d$ in this space) and as elements of the grid, using a
distance induced by the topological relation between them. We assume a
4-neighborhood in our grid, so that the grid neighbors of unit
$u^{i,j}$ are the four units $u^{i-1,j}, u^{i+1,j}, u^{i,j-1},
u^{i,j+1}$. The grid distance between units induced by this choice is
the so-called \emph{chemical distance}
\begin{equation}
  \delta(u^{i,j}, u^{i',j'}) = |i-i'| + |j-j'|
\end{equation}
The units also have, between themselves, a distance \emph{qua} points
in the word space.  Because of normalization, the units are placed on
the unit sphere of the word space, a topology that is well captured if
we use, rather than the restriction of a distance defined in the whole
space, the \emph{cosine similarity}:
\begin{equation}
  \label{whoopsie}
  s(u^{i,j},u^{i',j'}) = \sum_{i=1}^W u_i^{i,j}u_i^{i',j'}
\end{equation}
All weights are positive (viz.\ all units are in the first octant of
the unit sphere), therefore $s(u^{i,j},u^{i',j'})\in[0,1]$, and we can
define the word space distance between two units as
\begin{equation}
  d(u^{i,j},u^{i',j'}) = 1 - s(u^{i,j},u^{i',j'})
\end{equation}

The grid of units is placed in the word space following a learning
procedure based on documents that the user has considered interesting
in the past and which constitute our initial training set. Let
${\mathcal{D}}=\tset{D_1,\ldots,D_n}$ be such a set of documents. Each
document is processed by stop-word removal and stemming using standard
information retrieval techniques \cite{salton:88}. Using these
techniques, document $D_i$ is modeled as a \emph{bag} (multi-set) of
stems $D_i=\tbag{t_{i,1},\ldots,t_{i,n_i}}$. For each stem $t$ we
determine its \emph{document frequency}, defined simply as the number
of times that stem appears in the document (viz., in the bag)
$D_i$. We know from the information retrieval literature
\cite{salton:88} that the stem frequency is a poor indicator of the
relevance of a stem for the characterization of a document. Common
words will appear many times in any document, and will consequently be
of little help for discriminating between documents.

A good word for characterizing a document is a word that appears
many times in the document but is relatively rare in the English
language. We therefore consider a standard corpus of the English
language \cite{bnccorpus:93} and determine for each word its
\emph{inverse frequency}: if the corpus contains $C$ words, and the
stem $t$ appears $C_t$ times, then we define its \emph{raw frequency}
as
\begin{equation}
  r_t = \frac{C_t}{C}
\end{equation}
and its (logarithmic) \emph{inverse corpus frequency} as
\begin{equation}
  \mbox{icf}_t = \log \frac{1}{r_t} = - \log r_t = \log C - \log C_t
\end{equation}
(the presence of the logarithm is standard in information retrieval
and is based on empirical considerations). If the stem $t_j$ appears
$n_{ij}$ times in document $D_i$, and the document contains $|D_i|$
words, then the \emph{weight} of the stem is given by
\begin{equation}
  \label{weighting}
  q_{ij} = \frac{n_{ij}}{|D_i|} \mbox{icf}_{ij} = - \frac{n_{ij}}{|D_i|} \log r_{ij}
\end{equation}
These weights depend on the word frequency of the whole document but,
in order to obtain a finer representation, we break the document into
sentences (a sentence is defined simply as a sequence of words
terminated by a period, a comma, a colon or a semicolon) and represent
each one as a separate point using the words that appear in it and the
weights $q_{ij}$, computed on the whole document $D_i$. If the
sentence is composed of the stems $\tset{w_{k_1},\ldots,w_{k_s}}$,
then it is represented as the point of coordinates
\begin{equation}
  \label{sentence}
  q = (0,\ldots,0,q_{k_1},0,\ldots,q_{k_1},\ldots,0,q_{k_s},\ldots,0)
\end{equation}
Note again that a point is a representation of a sentence, but the
coordinates of the point are weights determined on the basis of the
document of which the sentence is part. It is therefore possible 
for the same sentence that appears in different documents to have
different representations. 

Finally, the coordinates of the sentence are normalized in order to
place the point on the unit sphere. The final representation of the
sentence is therefore the point
\begin{equation}
  p = \frac{1}{\left[ \sum_{i=1}^s q_{k_i}^2 \right]^{1/2}}
  (0,\ldots,0,q_{k_1},0,\ldots,q_{k_2},\ldots,0,q_{k_s},\ldots,0)
\end{equation}
Document $D_i$ is represented as a bag of points in the word space,
one for each sentence that appear in it
$S_i=\tbag{p_{i,1},\ldots,p_{i,k_i}}$. The collection ${\mathcal{D}}$
will be represented by the bag-union of its documents:
\begin{equation}
  {\mathcal{S}}=\bigcup\!\!\!\!\!\!\!+\,\,\, S_i
\end{equation}

During learning, the elements of the training set ${\mathcal{S}}$ are
presented one at a time (viz., one sentence presentation at a time).
Each presentation of an element (and the consequent learning
procedure, detailed below) is an \emph{event}; a presentation of all
the points in the set ${\mathcal{S}}$ is an \emph{epoch}. Learning
consists of a suitable number of epochs.

Upon presentation of a point $p$, the similarity with all units of the
grid is computed. The \emph{best matching unit} (BMU,
indicated as $u^*$) is the unit $u^{i,j}$ with the maximum similarity.
\begin{equation}
  \label{bamueq}
  u^* = \argmax_{i,j} s(p,u^{i,j}) = \argmax_{i,j} \sum_{i=1}^W p_i u_i^{i,j} 
\end{equation}
The idea behind the self-organizing map is that the BMU $u^*$ will be
moved by a certain amount towards the point $p$, and that the units in
a suitable neighborhood of $u^*$ (where the neighborhood is intended
in the grid topology) will be moved as well, by an extent decreasing
as the grid distance from the BMU increases.  Define the functions:
\begin{description}
\item[$\zeta(t)$:] learning factor at time $t$; the function $\zeta$,
  with $0\le\zeta\le1$ is monotonically non-increasing with $t$, and 
  \[
  \lim_{t\rightarrow\infty} \zeta(t) = 0
  \]
  thus guaranteeing the stability of the learning process. The
  decrease rate $\dot{\zeta}(t)<0$ must be small enough to guarantee a
  good quality of learning.
\item[$h(t,\delta)$:] the neighborhood function, which controls how
  much the units at distance $\delta$ from the winning unit will
  move. This function is such that:
  \begin{equation}
    \begin{aligned}
      \frac{\partial h}{\partial t} &\le 0 \\
      h(t,\delta) &\in [0,1]\ \ \mbox{(for all $t$, $\delta$)} \\
      h(t,\delta+1) &\le h(t,\delta)\ \ \mbox{(for all $t$)} \\
      h(t,0) &= 1 \ \ \mbox{(for all $t$)}
    \end{aligned}
  \end{equation}
\end{description}
with these definitions, upon the presentation of a point $p$, each
unit $u^{i,j}$ will be updated according to the rule:
\begin{equation}
  u^{i,j} \leftarrow u^{i,j} + \zeta(t) h(t,\delta(u^*,u^{i,j})) [p - u^{i,j}]
\end{equation}

In our test we used for $\zeta$ the function
\begin{equation}
  \label{zetaeq}
  \zeta(t) = \frac{1}{\phi_{u}+\gamma_{u}(t)+1}
\end{equation}
where $\phi_u$ is a confidence value for unit $u$, set to $1$ in our
test, and $\gamma_u(t)$ is the number of times unit $u$ has been the
BMU up to time $t$.  For $h(t,\delta)$ we use a mexican hat function
$h(t,\delta)=F_{1/t}(\delta)$, with
\begin{equation}
  \label{eqMexicanHat}
  F_\sigma (d) = \frac{2}{\sqrt{3}\sigma\pi ^{\frac{1}{4}} } (1-\frac{d^2}{\sigma^2})  e^{-\frac{d^2}{2\sigma^2}} 
\end{equation}
Compared with the Gaussian neighboring function used in previous work
\cite{santini:09a}, this neighborhood function produces a bigger
change in the units close to the BMU and will decrease till pushing
away units that are not close enough to the BMU. This repulsive
action at intermediate distances counters the tendency of the network
to concentrate too much in the dense areas of the input space,
resulting sometimes in a large number of nearly identical (and
therefore not very informative) units. During the learning process, the
radius of the neighborhood is decremented over time (by reducing
$\sigma$) so that the units that are far away will be less influenced.

When learning is done, the map is laid in the word space in a way that
approximates the probability distribution of the input points
(viz.\ the representation of the sentences) subject to the two
constraints of two-dimensionality and continuity. If the number of
units is sufficiently large, the map exhibit certain optimality
properties \cite{santini:96a}. This map constitutes what we call the
\emph{latent semantic manifold}, akin to the semantic subspace of
latent semantics \cite{deerwester:90}, but not constrained by the
requirement of being a linear sub-space of the word space.

\subsection{Information Filtering with the standard model}
Once the network has been trained, it constitutes the \emph{context}
that we use for filtering. Whenever a document arrives, it is assigned
a score using the following procedure:
\begin{description}
\item[i)] the document is processed using the standard techniques used
  for the documents in the training set, but without breaking it into
  sentences: weights are assigned using (\ref{weighting}) then
  normalized. The result is a representation of the document as a
  single point in the word space;
\item[ii)] the similarity with all units is computed, and the BMU is
  determined as in (\ref{bamueq});
\item[iii)] the BMU determines the semantic area of the document
  (viz.\ the part of the context for which it is relevant), and the
  similarity
  \begin{equation}
    \label{absolute}
    s_p = \sum_{k=1}^W p_k u_k^*
  \end{equation}
  determines the \emph{absolute relevance} of the document for the
  context.
\end{description}

As in the Robertsonian model, the most relevant items are selected for
display.

\subsection{Display of the Result}
Documents arrive in a stream, so we have to consider the presence of a
potentially infinite number of them, creating the issue of how to
display potentially infinite results in a finite display. There are
several methods that one can use to display at any given time a
significant set of documents. The details and the nature of such
methods depend on many variables, among which are the nature of the
application or the layout of the display
\cite{santini:12b,santini:18a}. A detailed study of such methods is
beyond the scope of this paper but, for the sake of concreteness, we
shall assume that the documents to be displayed are chosen as follows.

Assume that, at a given time $t$, the interface is showing the
documents $\tlist{d_1,\ldots,d_q}$, having relevances
$\tlist{s_1,\ldots,s_q}$. Assume also that, regardless of the order in
which the documents are actually shown in the interface, the list is
ordered in decreasing score values. At this time, document $d_p$
arrives, and the algorithm gives it a score $s_p$. Then:
\begin{description}
\item[i)] if $s_p<s_q$, the document $d_p$ is not shown;
\item[ii)] if $s_{k-1}\le{s_p}\le{s_k}$ the document $d_p$ is inserted
  in the $k$th position of the list, which now becomes
  \[
  \tlist{s_1,\ldots,s_{k-1},s_p,s_k,\ldots,s_{q-1}}
  \]
  document $d_q$ is no longer shown;
\item[iii)] the relevance of all document in the list is decreased by
  a factor $\beta<1$:
  \begin{equation}
    \label{redlst}
    s_k \leftarrow \beta s_k\ \ \ k=1,\ldots,q
  \end{equation}
\end{description}

Step iii) implements a \dqt{loss of relevance} of document that have
been already displayed for some time, and avoids the undesirable
situation in which very relevant documents stay forever in the list,
preventing new documents from being displayed.

\section{Novelty and Coverage}
The concept of novelty has become important in information retrieval
research as it may help to overcome some of the drawbacks of the
standard Robertsonian model \cite{robertson:76}, in which each
document is evaluated independently for relevance to a query or a
situation, and the most relevant documents are presented to the
user. The Robertsonian model is based on a number of specific (and
often unspoken) assumptions about relevance \shortcite{saracevic:07},
of which the one we are especially interested in is
\emph{independence}: \emph{the relevance of a document does not depend
  on the other documents of the result set}.  Starting towards the end
of the 1990s, these assumptions have been questioned
\cite{clarke:09,carbonell:98} and different, \emph{non-Robertsonian}
models of information retrieval have been proposed.

In teh case of the independence assumption, this questioning has
materialized in the the realization that the relevance of a document
is relative to the presence of other documents in the result set:
retrieving a document that, taken in isolation, would have been
relevant, might not be so relevant if the data set already
contains documents similar to it %
\footnote{This realization is, in an embryonic way, present in the
  Robertsonian model: if we stand by the purely formal statement of
  the model, the optimal result set is obtained by repeating the most
  relevant document as many times as necessary to fill the result
  list. So, even in the Robertsonian model, the hypothesis of
  independence must be suspended at least to the point of removing
  duplicates.}%
.

In a \dqt{good} list of documents, each document should be, to a
certain degree, \emph{novel}, in the sense that each should provide
information that the other documents of the list do not provide. We
are interested, let us say, in Marcel Proust. One of aspects of the
subject is the writer's biography. If, however, the first document in
the set is \dqt{the} ultimate biography of Marcel Proust, further
bibliographies will not add much information to the set. For the
second document in the set we will probably prefer something
different, for instance, something on Proust's novels. If the second
document is, say, a complete critical analysis of \emph{A la recherce
  du temps perdu}, we might want, as a third document, something on
\emph{Pastiches et m\'elanges}, or some study on the reception of
Proust's work, or even a reference to Alain de Botton's book \emph{How
  Proust can change your life}. It is not enough that all documents of
the list be about Proust: to attain a high novelty each one must be in
some respect unique.

Suppose that we have executed a query in an Information Retrieval
system and we have received as result the set of documents
$R=\{d_1,\ldots,d_n\}$.  The novelty of document $d_i$ with respect to
the set $R$ is the extent to which $d_i$ contains information that the
set $R\backslash\{d_i\}$ does not.  Novelty is therefore a property of
a document with respect to other documents in a set; we can measure
the overall novelty of a set by considering a suitable average of the
novelty of its elements. Standard measures of novelty and of the
related concept of diversity have been proposed and evaluated in
information retrieval and recommendation systems \cite{vargas:12}.

In this paper, we shall change somewhat this concept of novelty due to
the specific circumstances in which we work. Two characteristics of
the situation that we are studying require this change.

\begin{description}
\item[i)] In the standard concept, novelty is defined within the
  confines of a specific query: a person is interested in something
  rather specific (the subject of a query) and documents should be
  novel within the narrow straitjacket imposed by the query: in the
  case of the Marcel Proust query, a document about the mating habits
  of the woodpecker would certainly be very novel but, alas, it would
  also be completely irrelevant. Our straitjacket is looser: we
  develop a model of a person's general interests (what we call the
  \emph{context} of that person) and we filter information based on
  \emph{all} these interests: any piece of information on a subject
  the person is interested in will be valuable, and novelty has to be
  defined within these new, wider, limits.
\item[ii)] Standard novelty is defined for a result set, which is a
  one-shot affair: somebody makes a query (or asks for a
  recommendation) and receives a result. We are, on the other hand,
  concerned with a \emph{stream} of information coming continuously to
  us and coming to a person whose interests also evolve with time. The
  concept of novelty must take into account this time-dependence: a
  document that is not novel today might very well be novel tomorrow:
  if I am interested in world affairs and I just received a very good
  article on the result of the Presidential elections in Ecuador, new
  results about the same presidential elections might not be very
  novel \emph{today}. But if in a week a new article on the
  presidential elections appears, that probably means that there is
  something new about the topic, and I might be very interested in the
  article. That is, a piece of data that now is not novel due to the
  previous reception of similar data might be novel if we allow enough
  time to pass.
\end{description}

The standard concept of novelty describes simply the relation between
a document and a set of documents, since no other source is available
to determine whether a document is novel or not. In our case, however,
we do have additional information in the context model that we use to
determine the person's interests (viz., the \emph{context}), so it
seems natural to define an \emph{external} notion of novelty, no
longer intrinsic to the result data themselves, but relative to the
context. At the same time, we want our novelty model to be dynamic, so
that a document that is not novel at a certain time, may become novel
at a later time.  A conceptual definition of novelty, from our point
of view, could be the following

\begin{quote}
  A document is \emph{novel} it it belongs to a semantic area of
  interest to a person for which no documents have been seen in the
  recent past.
\end{quote}

We can make this concept more grounded by making reference to our user
model. We have seen that each unit in the model represents a
\dqt{semantic area} of interest to the user. We associate to each
semantic area (viz., to each unit) a value, called \emph{urgency} or
\emph{interest}: $\lambda^k(t)\in[0,1]$ is the urgency of semantic
area $k$ at time $t$. We shall analyze the evolution of $\lambda^k(t)$
in the next section but, broadly speaking, we have
$\lambda^k(t)\sim{0}$ if we just received an item about semantic area
$k$, we have $\lambda^k(t)\sim{1}$ if semantic area $k$ hasn't been
covered by items in a long time. With these concepts in mind, we can
give our qualitative definition of \emph{novelty} of a data item:

\begin{quote}
  An item is \emph{novel} if it cover a semantic area with high
  urgency.
\end{quote}

Note again that our definition is external to the result set: the
novelty of an item at time $t$ is not determined by reference to the
other items in the result set that are being displayed at time $t$,
but by the fact that the portion of the context which the item refers
to is, at the time of arrival of the item, considered interesting.

We have defined (qualitatively, so far), the external novelty of an
item. Extending it to a cumulative measure for a set of items is not
straightforward, as the cumulative value will depend, in general, on
the order in which the items arrive and the arrival times. To see why
this is the case, consider a situation in which the same item is seen
in input twice with no intervening items. In this case, the joint
novelty will be basically that of the first occurrence since, after
the arrival of the first item, the interest of the semantic area to
which both belong will be virtually zero. However, if the second item
arrives a long time after the first, the urgency of that semantic area
will have recovered, so that both items will be considered to be
novel.

So, in order to have a cumulative definition of novelty, we need to
know not only what elements have we received but also their order and
the times of their arrival:

\begin{quote}
  A list of items $[w_1,\ldots,w_n]$ arriving at times
  $[t_1,\ldots,t_n]$ with no intervening items have high
  \emph{novelty} if each one is close to a unit that, at the time of
  arrival, has high urgency.
\end{quote}

This measure is very specific and ill-suited for considerations and
experiments of a general nature. In order to get a more manageable
measure, we define \emph{coverage}, a measure that abstracts from
order and arrival times, and simply counts the number os semantic
areas that are activated:

\begin{quote}
  A set of items $\{w_1,\ldots,w_n\}$ has high \emph{coverage} if each
  one of the item is close to a different semantic unit.
\end{quote}

We shall make these measures specific, and give quantitative
definitions in the next section.

\section{Coverage-enforcing algorithm}
\label{noveltySOM}
Our goal in this section is to devise algorithms to increase \emph{coverage}, that is, to favor, among the
items that arrive in a given time span, those that cover parts of the
user's semantic field that had not been covered in the recent
past. We want to strike a balance in doing this: we don't
want to display a marginally relevant document at the expense of a
highly relevant one only because the former is weakly related to an
area in the semantic field that hasn't been seen in a while. 

In order to strike this balance, we introduce, for each unit of the
network, a time-varying parameter, which we have previously called the
\emph{urgency} or \emph{interest}, $\lambda^{ij}\in[0,1]$. The value
of $\lambda^{ij}$ defines how much, at time $t$, the user is
interested in receiving items in the semantic area of unit
$u^{ij}$. Initially, the value of $\lambda^{ij}$ is set to 1 for all
units. If at time $t$ unit $u^{ij}$ is the BMU and the item that just
arrived is displayed, then we assume that the user is no longer
interested in the semantic area of $u^{ij}$, and $\lambda^{ij}$ is
reduced accordingly. Each time $u^{ij}$ is not the BMU or whenever,
$u^{ij}$ being the BMU, the item is not displayed, we assume that the
user \dqt{forgets} having seen an item in that semantic area, and that
his interest in it increases until, after a certain time, it is
restored to its initial value. The specific form of this variation
doesn't seem to be important, so we choose the simplest: a fractional
drop in interest when the unit is the BMU, followed by a linear
recovery. Let $\Theta$ be the drop constant (viz., the factor
by which $\lambda^{ij}$ is reduced when $u^{ij}$ is the BMU), and $K$
the recovery constant (the time it takes for $\lambda^{ij}$ to get
back to its original value). Then
\begin{equation}
  \lambda^{ij}(t+1) = 
  \begin{cases}
    \displaystyle \frac{\lambda^{ij}(t)}{\Theta} & \mbox{if $u^{ij}$ is the BMU and the item is displayed} \\
    \displaystyle \min\Bigl\{ 1, \lambda^{ij}(t) + \frac{\Theta-1}{K\Theta}\Bigr\} & \mbox{otherwise}
  \end{cases}
\end{equation}
We call this solution the \emph{drastic} interest update: the interest
of the BMU is decreased, while that of all other units is increased
or kept constant at 1. 
\def\csize{0.5}
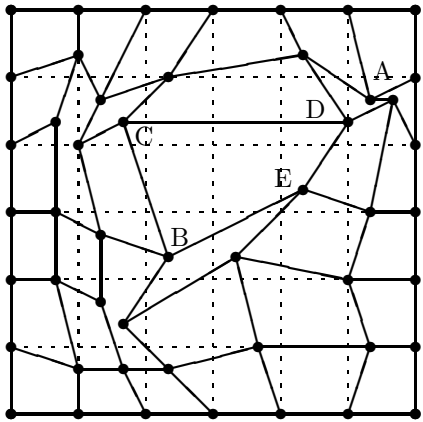
\begin{figure}
  \setlength{\unitlength}{0.85em}
  \begin{center}
    \begin{picture}(21,21)(0,0)
     \multiput(0,0)(0,3){7}{%
       \multiput(0,0)(0.6,0){30}{%
         \line(1,0){0.15}%
       }
     }
     \multiput(0,0)(3,0){7}{%
        \multiput(0,0)(0,0.6){30}{%
          \line(0,1){0.15}%
        }
      }
      \thicklines
      \multiput(0,0)(18,0){2}{%
        \multiput(0,0)(0,3){7}{%
          \put(0,0){\circle*{\csize}}%
        }
      }
      \multiput(0,0)(0,18){2}{%
        \multiput(0,0)(3,0){7}{%
          \put(0,0){\circle*{\csize}}%
        }
      }
     \multiput(0,0)(0,18){2}{\line(1,0){18}}
     \multiput(0,0)(18,0){2}{\line(0,1){18}}
     \put(0,3){\line(3,-1){3}}
     \put(3,2){\circle*{\csize}}
     \put(3,2){\line(1,0){2}}
     \put(3,0){\line(0,1){2}}
     \put(5,2){\circle*{\csize}}
     \put(5,2){\line(1,0){2}}
     \put(6,0){\line(-1,2){1}}
     \put(7,2){\circle*{\csize}}
     \put(7,2){\line(4,1){4}}
     \put(9,0){\line(-1,1){2}}
     \put(11,3){\circle*{\csize}}
     \put(11,3){\line(1,0){5}}
     \put(12,0){\line(-1,3){1}}
     \put(16,3){\circle*{\csize}}
     \put(16,3){\line(1,0){2}}
     \put(15,0){\line(1,3){1}}
     \put(3,2){\line(-1,4){1}}
     \put(0,6){\line(1,0){2}}
     \put(2,6){\circle*{\csize}}
     \put(5,2){\line(-1,3){1}}
     \put(2,6){\line(2,-1){2}}
     \put(4,5){\circle*{\csize}}
     \put(7,2){\line(-1,1){2}}
     \put(4,5){\line(1,-1){1}}
     \put(5,4){\circle*{\csize}}
     \put(11,3){\line(-1,4){1}}
     \put(5,4){\line(5,3){5}}
     \put(10,7){\circle*{\csize}}
     \put(16,3){\line(-1,3){1}}
     \put(10,7){\line(5,-1){5}}
     \put(15,6){\circle*{\csize}}
     \put(15,6){\line(1,0){3}}
     \put(0,9){\line(1,0){2}}
     \put(2,6){\line(0,1){3}}
     \put(2,9){\circle*{\csize}}
     \put(2,9){\line(2,-1){2}}
     \put(4,5){\line(0,1){3}}
     \put(4,8){\circle*{\csize}}
     \put(4,8){\line(3,-1){3}}
     \put(5,4){\line(2,3){2}}
     \put(7,7){\circle*{\csize}}
     \put(7.1,7.5){\makebox(0,0)[bl]{B}}
     \put(7,7){\line(2,1){6}}
     \put(10,7){\line(1,1){3}}
     \put(13,10){\circle*{\csize}}
     \put(12.5,10.1){\makebox(0,0)[rb]{E}}
     \put(13,10){\line(3,-1){3}}
     \put(15,6){\line(1,3){1}}
     \put(16,9){\circle*{\csize}}
     \put(16,9){\line(1,0){2}}
     \put(0,12){\line(2,1){2}}
     \put(2,9){\line(0,1){4}}
     \put(2,13){\circle*{\csize}}
     \put(2,13){\line(1,-1){1}}
     \put(4,8){\line(-1,4){1}}
     \put(3,12){\circle*{\csize}}
     \put(3,12){\line(2,1){2}}
     \put(7,7){\line(-1,3){2}}
     \put(5,13){\circle*{\csize}}
     \put(5.5,12.8){\makebox(0,0)[lt]{C}}
     \put(5,13){\line(1,0){10}}
     \put(13,10){\line(2,3){2}}
     \put(15,13){\circle*{\csize}}
     \put(14.0,13.2){\makebox(0,0)[rb]{D}}
     \put(15,13){\line(2,1){2}}
     \put(16,9){\line(1,5){1}}
     \put(17,14){\circle*{\csize}}
     \put(17,14){\line(1,-2){1}}
     \put(0,15){\line(3,1){3}}
     \put(2,13){\line(1,3){1}}
     \put(3,16){\circle*{\csize}}
     \put(3,16){\line(1,-2){1}}
     \put(3,12){\line(1,2){1}}
     \put(4,14){\circle*{\csize}}
     \put(4,14){\line(3,1){3}}
     \put(5,13){\line(1,1){2}}
     \put(7,15){\circle*{\csize}}
     \put(7,15){\line(6,1){6}}
     \put(15,13){\line(-2,3){2}}
     \put(13,16){\circle*{\csize}}
     \put(13,16){\line(3,-2){3}}
     \put(17,14){\line(-1,0){1}}
     \put(16,14){\circle*{\csize}}
     \put(16.1,14.9){\makebox(0,0)[lb]{A}}
     \put(16,14){\line(2,1){2}}
     \put(3,18){\line(0,-1){2}}
     \put(6,18){\line(-1,-2){2}}
     \put(9,18){\line(-2,-3){2}}
     \put(12,18){\line(1,-2){1}}
     \put(15,18){\line(1,-4){1}}
    \end{picture}
  \end{center}
  \caption{\capstyle A non-uniform network with units concentrated in certain
    areas of the word space, resulting from a skewed collection of
    context items. The result is areas of low density of units (the
    center, in this case) together with areas of high density. For the
    meaning of the labels \dqt{A,B,C,D,E} in this figure, see the
    text.}
  \label{concentrated}
\end{figure}

The algorithm that we use to create the context guarantees that nearby
units on the grid are also relatively close in the word space, that
is, that there is a certain degree of semantic overlap between
them. Because of this, it is not unreasonable to assume that when the
interest in the BMU drops, the interest in nearby units should also
drop, albeit to a lesser degree. In order to implement this
observation, we define an action radius $\Delta\ge{0}$ and a grading
function
\begin{equation}
  \rho(\delta) = 1 - \frac{\delta}{\Delta}
\end{equation}

The interest factor is then updated for all units according to
\begin{equation}
  \label{verhor}
  \lambda^{ij}(t+1) = 
  \begin{cases}
    \displaystyle \Bigl[1 + \bigl( \frac{1}{\Theta}-1 \bigr) \rho(\delta(u^{ij},u^*)) \Bigr] \lambda^{ij}(t)
     & \mbox{if $\delta(u^{ij},u^*)\le\Delta$} \\
    \displaystyle \min\Bigl\{ 1, \lambda^{ij}(t) + \frac{\Theta-1}{K\Theta}\Bigr\} & \mbox{otherwise}
  \end{cases}
\end{equation}
Units whose distance from the BMU is less than $\Delta$ will have
their interest reduced by an amount that will be smaller the farther
away the unit is from the BMU. Unit at a distance $\Delta$ from the
BMU will have their interest unchanged, while units at a distance
greater than $\Delta$ will have their interest increased or kept
constant to 1. We call this solution the \emph{graded} interest
update.

The values $\lambda^{ij}$ are used to modify the equation that
determines the BMU, penalizing units with lower interest. Equation
(\ref{bamueq}) is replaced by
\begin{equation}
  \label{blonk}
  u^* = \argmax_{i,j} \lambda^{ij} s(p,u^{i,j}) = \argmax_{i,j} \lambda^{ij} \sum_{k=1}^W p_k u_k^{i,j} 
\end{equation}
We call the term $\lambda^{ij}\sum_{k=1}^W{p_k}u_k^{i,j}$ the
\emph{modulated relevance}, distinguishing it from the absolute
relevance defined in (\ref{absolute}). This equation penalizes units
with lower interest so that if an item in the semantic area of unit
$u^{ij}$ has been displayed recently, new items in the same semantic
areas will have a lower modulated relevance, and will need a higher
absolute relevance in order to be selected, making items in other
semantic areas more likely to be displayed.
\def\csize{0.5}
\begin{figure}[bhtp]
  \setlength{\unitlength}{0.85em}
  \begin{center}
    \begin{picture}(21,21)(0,0)
     \multiput(0,0)(0,3){7}{%
       \multiput(0,0)(0.6,0){30}{%
         \line(1,0){0.15}%
       }
     }
     \multiput(0,0)(3,0){7}{%
        \multiput(0,0)(0,0.6){30}{%
          \line(0,1){0.15}%
        }
      }
      \thicklines
      \multiput(0,0)(18,0){2}{%
        \multiput(0,0)(0,3){7}{%
          \put(0,0){\circle*{\csize}}%
        }
      }
      \multiput(0,0)(0,18){2}{%
        \multiput(0,0)(3,0){7}{%
          \put(0,0){\circle*{\csize}}%
        }
      }
     \multiput(0,0)(0,18){2}{\line(1,0){18}}
     \multiput(0,0)(18,0){2}{\line(0,1){18}}
     \put(0,3){\line(3,-1){3}}
     \put(3,2){\circle*{\csize}}
     \put(3,2){\line(1,0){2}}
     \put(3,0){\line(0,1){2}}
     \put(5,2){\circle*{\csize}}
     \put(5,2){\line(1,0){2}}
     \put(6,0){\line(-1,2){1}}
     \put(7,2){\circle*{\csize}}
     \put(7,2){\line(4,1){4}}
     \put(9,0){\line(-1,1){2}}
     \put(11,3){\circle*{\csize}}
     \put(11,3){\line(1,0){5}}
     \put(12,0){\line(-1,3){1}}
     \put(16,3){\circle*{\csize}}
     \put(16,3){\line(1,0){2}}
     \put(15,0){\line(1,3){1}}
     \put(3,2){\line(-1,4){1}}
     \put(0,6){\line(1,0){2}}
     \put(2,6){\circle*{\csize}}
     \put(5,2){\line(-1,3){1}}
     \put(2,6){\line(2,-1){2}}
     \put(4,5){\circle*{\csize}}
     \put(7,2){\line(-1,1){2}}
     \put(4,5){\line(1,-1){1}}
     \put(5,4){\circle*{\csize}}
     \put(11,3){\line(-1,4){1}}
     \put(5,4){\line(5,3){5}}
     \put(10,7){\circle*{\csize}}
     \put(16,3){\line(-1,3){1}}
     \put(10,7){\line(5,-1){5}}
     \put(15,6){\circle*{\csize}}
     \put(15,6){\line(1,0){3}}
     \put(0,9){\line(1,0){2}}
     \put(2,6){\line(0,1){3}}
     \put(2,9){\circle*{\csize}}
     \put(2,9){\line(2,-1){2}}
     \put(4,5){\line(0,1){3}}
     \put(4,8){\circle*{\csize}}
     \put(4,8){\line(3,-1){3}}
     \put(5,4){\line(2,3){2}}
     \put(7,7){\circle*{\csize}}
     \put(7.1,7.5){\makebox(0,0)[bl]{B}}
     \put(7,7){\line(2,1){6}}
     \put(10,7){\line(1,1){3}}
     \put(13,10){\circle*{\csize}}
     \put(12.5,10.1){\makebox(0,0)[rb]{E}}
     \put(13,10){\line(3,-1){3}}
     \put(15,6){\line(1,3){1}}
     \put(16,9){\circle*{\csize}}
     \put(16,9){\line(1,0){2}}
     \put(0,12){\line(2,1){2}}
     \put(2,9){\line(0,1){4}}
     \put(2,13){\circle*{\csize}}
     \put(2,13){\line(1,-1){1}}
     \put(4,8){\line(-1,4){1}}
     \put(3,12){\circle*{\csize}}
     \put(3,12){\line(2,1){2}}
     \put(7,7){\line(-1,3){2}}
     \put(5,13){\circle*{\csize}}
     \put(5.5,12.8){\makebox(0,0)[lt]{C}}
     \put(5,13){\line(1,0){10}}
     \put(13,10){\line(2,3){2}}
     \put(15,13){\circle*{\csize}}
     \put(14.0,13.2){\makebox(0,0)[rb]{D}}
     \put(15,13){\line(2,1){2}}
     \put(16,9){\line(1,5){1}}
     \put(17,14){\circle*{\csize}}
     \put(17,14){\line(1,-2){1}}
     \put(0,15){\line(3,1){3}}
     \put(2,13){\line(1,3){1}}
     \put(3,16){\circle*{\csize}}
     \put(3,16){\line(1,-2){1}}
     \put(3,12){\line(1,2){1}}
     \put(4,14){\circle*{\csize}}
     \put(4,14){\line(3,1){3}}
     \put(5,13){\line(1,1){2}}
     \put(7,15){\circle*{\csize}}
     \put(7,15){\line(6,1){6}}
     \put(15,13){\line(-2,3){2}}
     \put(13,16){\circle*{\csize}}
%
%
%
     \put(3,18){\line(0,-1){2}}
     \put(6,18){\line(-1,-2){2}}
     \put(9,18){\line(-2,-3){2}}
     \put(12,18){\line(1,-2){1}}
%
     \put(10,8.5){\circle*{\csize}}
     \put(10,8.5){\line(0,1){4.5}}
     \put(10,13){\circle*{\csize}}
     \put(6,10){\circle*{\csize}}
     \put(10,10.75){\circle*{\csize}}
     \put(10.4,10.75){\makebox(0,0)[l]{F}}
     \put(6,10){\line(6,1){4}}
    \end{picture}
  \end{center}
  \caption{\capstyle The same network after inserting and removing units. New
    units have been created in the mid-point of the segments $B$-$C$,
    $C$-$D$, and $B$-$E$, while unit $A$, placed in a zone of high
    density, has been removed. If necessary, the point marked F,
    mid-point of the line that joins two of the newly created units,
    will also be created.}
  \label{equalized}
\end{figure}
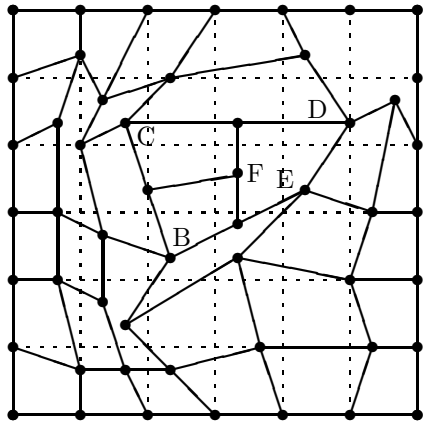

\subsection{Dynamic network topology}
In the standard WEBSOM algorithm, the topology of the network is
fixed: its shape (a regular rectangular grid) and the number of units
are set a priori. The analysis of early tests convinced us that we
needed a more flexible architecture. It is common for people to have
certain dominating interests, and their context will contain many
documents related to these interests. Consequently, when the network
is deployed in the word space, some areas will show a high
concentration of units. A schematic example, with a two-dimensional
word space is shown in Figure~\ref{concentrated}.
In this case, the areas near the corners of the square are the most
represented in the documents that formed the context and are,
consequently, the areas with the greatest concentration of units
\cite{kohonen:90}. This is what the network is supposed to do and,
were it not for the issue of novelty, it would be the optimal
configuration for the network to be in: it can be shown that, in the
limit of a continuum of units, this is the optimal two-dimensional
representation of the probability distribution from which the data are
drawn \cite{santini:96a}.

If we are interested in novelty, the situation of
Figure~\ref{concentrated} might not be optimal: novelty imposes a more
thorough exploration of the semantic space, even at the cost of
overemphasizing areas that are not so interesting but that haven't
been seen in a while. This requires a compromise between faithful
representation of the probability distribution of the training set and
coverage of the whole semantic space: while we still want the more
interesting areas to be more densely populated, we want to mitigate
this effect somewhat, so that no area is too dense or too sparse.  We
have modified the algorithm to allow a limited dynamic configuration
of the network, to make it more uniform than the raw algorithm would
do. Consider again the network in Figure~\ref{concentrated}. Here we
have a sparse center and a very dense upper-right corner. The sparsity
of the center can be identified from the fact that units $C$ and $D$
or $C$ and $B$ are very far away from each other in the word space,
despite being neighbor in the network topology. The opposite happens
around unit $A$. In order to mitigate this skew, our algorithm
will create three new units in the midpoints of the segments $C$-$D$,
$C$-$B$, and $B$-$E$ (units $E$ and $D$ are close enough that no new
unit is necessary). Note that in this way the network distance between
the pairs $(C,D)$, $(C,D)$ and $(B,E)$ has increased from one to
two. Additionally, we create the unit $F$ halfway between the new
units. In the upper-right corner, on the other hand, we reduce the
density by eliminating unit $A$. The resulting network topology is
that of Figure~\ref{equalized}.
This process is subject to a compromise: if we carried out creation
and deletion of unit to its extreme point, we would obtain a network
laid out uniformly in the word space, that is, te context would no
longer reflect the probability distribution of the training test. This
will give us maximum novelty, but will negatively impact the
precision.  Based on these considerations, we limit the number of new
units that can be created: each group of four units in the original
map will allow the creation of five new units, as in
Figure~\ref{newunit}. At the same time, no unit can be eliminated if
one of its original neighbors has already been eliminated.
\def\csize{0.5}
\begin{figure}[bthp]
  \setlength{\unitlength}{2em}
  \begin{center}
    \begin{picture}(6,5.5)(0,0)
      \newsavebox{\cross}
      \savebox{\cross}{
        \put(-0.25,-0.25){\line(1,1){0.5}}
        \put(-0.25,0.25){\line(1,-1){0.5}}
      }
      \thicklines
      \multiput(0,1)(0,4){2}{\line(1,0){6}}
      \multiput(1,0)(4,0){2}{\line(0,1){6}}
      \multiput(1,1)(4,0){2}{%
        \multiput(0,0)(0,4){2}{\circle*{\csize}}%
      }
      \thinlines
      \put(1,1){
        \put(-1,2){\line(1,0){6}}
        \put(2,-1){\line(0,1){6}}
        \put(2,0){\usebox{\cross}}
        \put(2,4){\usebox{\cross}}
        \put(0,2){\usebox{\cross}}
        \put(4,2){\usebox{\cross}}
        \put(2,2){\usebox{\cross}}
      }
    \end{picture}
  \end{center}
  \caption{\capstyle The creation of new units must be limited in order to avoid
    that the map become too uniform, thereby losing the information
    provided by the distribution of documents in the context. We only
    allow five new units to be created for each group of four original
    units. Here, the circles represent the original units of the maps,
    and the crosses represent the places where new units can be
    created.}
  \label{newunit}
\end{figure}
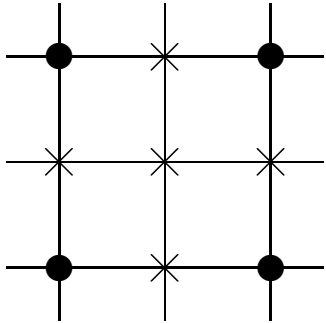

In order to implement this modification, we consider the following
quantities:

\begin{description}
\item[The \textbf{context diameter.}] The context diameter is the
  maximum distance between two units in the context, where the
  distance is defined in terms of the similarity computed as in
  (\ref{whoopsie}):
  \begin{equation}
    \label{eqActivationDist}
    d_{\phi} = \max\tset{1-s(u, v) | u,v \in \som}
  \end{equation}
  Note that, in the actual implementation, we estimate this distance
  by using a subset of the context documents, for the sake of efficiency.
\begin{table}[bht]
  \begin{center}
    \begin{tabular}{|l|p{4in}|}
      \hline
      $\som = \tset{u^{i,j} | 1 \le i,j \le N_s}$ & the network to be trained. \\
      \hline
      $S = \tset{p_1,\ldots,p_K}$ & the points that represent the sentences in the documents of the context. \\
      \hline
      $\zeta(t)$ & the instantaneous learning rate \\
      \hline
      $h(t,\delta)$ & the instantaneous neighborhood function \\
      \hline
      $d_{+}$ & the activation distance \\
      \hline
      $d_{-}$ & the deactivation distance \\
      \hline
      $nE$   & the maximum number of epochs \\
      \hline
      stop & the stop criterion function \\
      \hline
      active$[u]$ & attribute of each unit determining whether the unit is active.\\
      \hline
    \end{tabular}
  \end{center}
  \caption{\capstyle Input elements for the training algorithm.}
  \label{training_input}
\end{table}

\item[The \textbf{maximum activation distance}] is the maximum
  distance allowed between two nearby units before a new unit is
  created between them:
  \begin{equation}
    \label{eqMaxActivationDist}
    d_{+} = \frac{d_{\phi}}{\mu}
  \end{equation}
  where $\mu$ is a scaling constant. If two nearby units, say
  $u^{i,j}$ and $u^{i+1,j}$ (where \dqt{nearby} is to be interpreted
  in the 4-neighborhood topology), are, after a learning step, at a
  distance $d>d_{+}$, and and both them are from the initial grid,
  then a new unit $u^{i+1/2,j}$ is created between the two, in the
  mid-point of the line joining the two units:
  \begin{equation}
    \label{eqInitWeightsActivation}
    u_{k}^{i+1/2,j} = \frac{u_k^{i,j}+u_k^{i+1,j}}{2}, k=1,\ldots,W
  \end{equation}
\item[The \textbf{minimum deactivation distance}] is the minimum
  distance at which two unit can be before one of them is removed:
  \begin{equation}
    \label{eqMinDeactivationDist}
    d_{-} = \frac{d_{-}}{\nu}
  \end{equation}

  where $\nu>\mu$ is a scaling constant. If a unit has \emph{all}
  neighbors within a distance $d_{-}$, that unit is deactivated.
\end{description}

\def\csize{0.4}
\def\tsize{0.3}
\def\qsize{0.15}
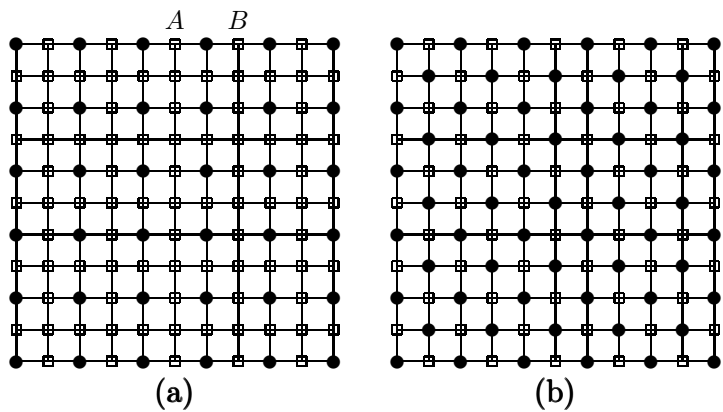
\begin{figure}[bhtp]
  \setlength{\unitlength}{1.2em}
  \begin{center}
    \begin{picture}(24,12)(0,-1)
      \newsavebox{\eunit}
      \savebox{\eunit}{
        \multiput(-\qsize,-\qsize)(\tsize,0){2}{\line(0,1){\tsize}}
        \multiput(-\qsize,-\qsize)(0,\tsize){2}{\line(1,0){\tsize}}
      }
      \put(0,0){
        \put(5,10.5){\makebox(0,0)[b]{$A$}}
        \put(7,10.5){\makebox(0,0)[b]{$B$}}
        \put(5,-0.5){\makebox(0,0)[t]{{\large \textbf{(a)}}}}
        \multiput(0,0)(0,1){11}{%
          \line(1,0){10}
        }
        \multiput(0,0)(1,0){11}{%
          \line(0,1){10}
        }
        \multiput(0,0)(2,0){6}{%
          \multiput(0,0)(0,2){6}{\circle*{\csize}}
        }
        \multiput(1,0)(2,0){5}{%
          \multiput(0,0)(0,1){11}{\usebox{\eunit}}
        }
        \multiput(0,1)(2,0){6}{%
          \multiput(0,0)(0,2){5}{\usebox{\eunit}}
        }
      }
      \put(12,0){
        \put(5,-0.5){\makebox(0,0)[t]{{\large \textbf{(b)}}}}
        \multiput(0,0)(0,1){11}{%
          \line(1,0){10}
        }
        \multiput(0,0)(1,0){11}{%
          \line(0,1){10}
        }
        \multiput(0,0)(0,2){6}{%
          \multiput(0,0)(2,0){6}{\circle*{\csize}}
        }
        \multiput(1,1)(0,2){5}{%
          \multiput(0,0)(2,0){5}{\circle*{\csize}}
        }
        \multiput(0,1)(0,2){5}{%
          \multiput(0,0)(2,0){6}{\usebox{\eunit}}
        }
        \multiput(1,0)(0,2){6}{%
          \multiput(0,0)(2,0){5}{\usebox{\eunit}}
        }
      }
    \end{picture}
  \end{center}
  \caption{\capstyle Two different organizations of the initial grid to allow a
    limited adaptation of the topology. The filled circles represent
    initially active units, the hollow squares initially inactive
    ones. The pictures represent the situation before learning is
    started. The interwoven network (a) has a lower ratio of active
    vs/inactive units, which entails a higher capacity for growth,
    while the checkerboard initialization (b) has an (approximately)
    equal number of active and inactive units.  Which solution gives
    better results depend in part on the characteristics of the
    context: in general, the checkerboard tends to perform better, as
    the low active/inactive ratio of the intertwined network often
    leads to relatively uniform networks in the word space, with
    consequent loss of precision. The network distance function is
    modified accordingly, so that only active units are counted. For
    example, units $A$ and $B$ in (a), which are separated only
    by one inactive unit, have $\delta(A,B)=1$.}
  \label{initialization}
\end{figure}
This limitation on the number of units that can be inserted allows us
an easy implementation. We set up a map on a grid of higher resolution
than actually needed and, at the beginning, leave some of the units
\emph{deactivated}. If two neighboring units get too far away from
each other (viz., at a distance greater than $d_+$) and there is a
deactivated unit between them, we switch it to \emph{activated} and
set its weights as in (\ref{eqInitWeightsActivation}). If two units
get too close to one another (viz., at a distance less than $d_=$), we
mark one of them as inactive. The network distance function is
modified accordingly so that only active units are counted. For
example, units $A$ and $B$ in Figure~\ref{initialization}.a, which are
separated only by one inactive unit, will have $\delta(A,B)=1$.

This solution allowed us to experiment with several initial
configurations of the network. Two such configurations are shown in
Figure~\ref{initialization}; we refer to them as the \emph{interwoven}
(on the left) and the \emph{checkerboard} (on the right)
configurations. The interwoven network has a lower ratio of active
vs/inactive units (there are approximately three inactive units for
each active one), which entails a higher capacity for growth, while
the checkerboard initialization has an (approximately) equal number of
active and inactive units.


\subsection{Algorithm summary}
In the previous sections we have presented the basic WEBSOM algorithm
and then we have defined a series of modifications. This has resulted
in a somewhat piecemeal presentation. To make the description clearer,
in this section we provide a brief summary of the final algorithms as
they result from the modifications of the previous section, as well as
the values of the parameters that we have used in the tests of the
following section.

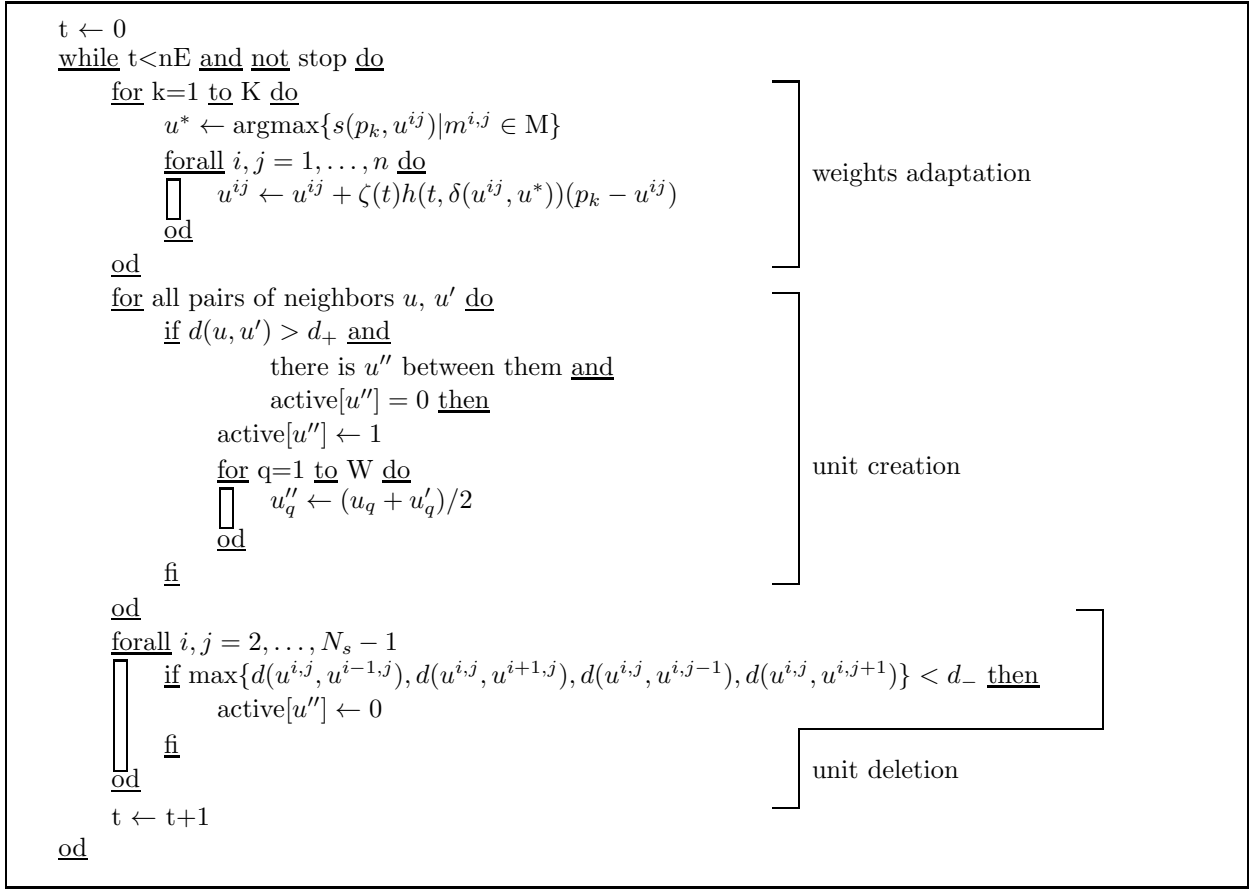
\begin{figure}[htbp]
  \begin{center}
    \setcounter{spc}{0}
    \setcounter{ind}{0}
    \def\istp{20}
    \def\vspc{13}
    \addtocounter{spc}{\vspc}
    \setlength{\unitlength}{0.1em}
    \begin{picture}(470,335)(0,-335)
      \multiput(0,-335)(470,0){2}{\line(0,1){335}}
      \multiput(0,-335)(0,335){2}{\line(1,0){470}}
      \put(20,0){
        \put(280,-30){\line(0,-1){70}}
        \put(280,-30){\line(-1,0){10}}
        \put(280,-100){\line(-1,0){10}}
        \put(285,-65){\makebox(0,0)[l]{weights adaptation}}
        \put(280,-110){\line(0,-1){110}}
        \put(280,-220){\line(-1,0){10}}
        \put(280,-110){\line(-1,0){10}}
        \put(285,-175){\makebox(0,0)[l]{unit creation}}
        \put(395,-230){\line(-1,0){10}}
        \put(395,-230){\line(0,-1){45}}
        \put(395,-275){\line(-1,0){115}}
        \put(280,-275){\line(0,-1){30}}
        \put(280,-305){\line(-1,0){10}}
        \put(285,-290){\makebox(0,0)[l]{unit deletion}}
        
        \put(\theind,-\thespc){\makebox(0,0)[lb]{t $\leftarrow$ 0}}
        \addtocounter{spc}{\vspc}
        \put(\theind,-\thespc){%
          \makebox(0,0)[lb]{\cmd{while} t$<$nE \cmd{and} \cmd{not} stop \cmd{do}}
        }
        \addtocounter{spc}{\vspc}
        \addtocounter{ind}{\istp}
        \put(\theind,-\thespc){%
          \makebox(0,0)[lb]{\cmd{for} k=1 \cmd{to} K \cmd{do}}
        }
        \addtocounter{spc}{\vspc}
        \addtocounter{ind}{\istp}
        \put(\theind,-\thespc){\makebox(0,0)[lb]{$u^* \leftarrow \argmax \tset{s(p_k, u^{ij}) | m^{i,j}\in\mbox{\som}}$}}
        \addtocounter{spc}{\vspc}
        \put(\theind,-\thespc){\makebox(0,0)[lb]{%
            \cmd{forall} $i,j = 1,\ldots,n$ \cmd{do}}%
            \multiput(1,-2)(5,0){2}{\line(0,-1){15}}%
            \multiput(1,-2)(0,-15){2}{\line(1,0){5}}
        }
        \addtocounter{spc}{\vspc}
        \addtocounter{ind}{\istp}
        \put(\theind,-\thespc){\makebox(0,0)[lb]{%
            $u^{ij}\leftarrow u^{ij} + \zeta(t)h(t, \delta(u^{ij},u^*))(p_k - u^{ij})$
          }
        }
        \addtocounter{spc}{\vspc}
        \addtocounter{ind}{-\istp} 
        \put(\theind,-\thespc){\makebox(0,0)[lb]{\cmd{od}}}
        \addtocounter{spc}{\vspc}
        \addtocounter{ind}{-\istp} 
        \put(\theind,-\thespc){\makebox(0,0)[lb]{\cmd{od}}}
        \addtocounter{spc}{\vspc}
        \put(\theind,-\thespc){\makebox(0,0)[lb]{\cmd{for} all pairs of neighbors $u$, $u'$ \cmd{do}}}
        \addtocounter{spc}{\vspc}
        \addtocounter{ind}{\istp} 
        \put(\theind,-\thespc){\makebox(0,0)[lb]{\cmd{if} $d(u,u')>d_+$ \cmd{and}}}
        \addtocounter{spc}{\vspc}
        \addtocounter{ind}{40} 
        \put(\theind,-\thespc){\makebox(0,0)[lb]{there is $u''$ between them \cmd{and}}}
        \addtocounter{spc}{\vspc}
        \put(\theind,-\thespc){\makebox(0,0)[lb]{active$[u'']=0$ \cmd{then}}}
        \addtocounter{ind}{-40} 
        \addtocounter{spc}{\vspc}
        \addtocounter{ind}{\istp} 
        \put(\theind,-\thespc){\makebox(0,0)[lb]{active$[u'']\leftarrow 1$}}
        \addtocounter{spc}{\vspc}
        \put(\theind,-\thespc){\makebox(0,0)[lb]{%
            \cmd{for} q=1 \cmd{to} W \cmd{do}}
            \multiput(1,-2)(5,0){2}{\line(0,-1){15}}%
            \multiput(1,-2)(0,-15){2}{\line(1,0){5}}
        }
        \addtocounter{spc}{\vspc}
        \addtocounter{ind}{\istp} 
        \put(\theind,-\thespc){\makebox(0,0)[lb]{$u_q^{\prime\prime} \leftarrow (u_q + u_q^{\prime})/2$}}
        \addtocounter{spc}{\vspc}
        \addtocounter{ind}{-\istp} 
        \put(\theind,-\thespc){\makebox(0,0)[lb]{\cmd{od}}}
        \addtocounter{spc}{\vspc}
        \addtocounter{ind}{-\istp} 
        \put(\theind,-\thespc){\makebox(0,0)[lb]{\cmd{fi}}}
        \addtocounter{spc}{\vspc}
        \addtocounter{ind}{-\istp} 
        \put(\theind,-\thespc){\makebox(0,0)[lb]{\cmd{od}}}
        \addtocounter{spc}{\vspc}
        \put(\theind,-\thespc){\makebox(0,0)[lb]{%
            \cmd{forall} $i,j=2,\ldots,N_s-1$}
            \multiput(1,-2)(5,0){2}{\line(0,-1){42}}%
            \multiput(1,-2)(0,-42){2}{\line(1,0){5}}
        }
        \addtocounter{spc}{\vspc}
        \addtocounter{ind}{\istp} 
        \put(\theind,-\thespc){\makebox(0,0)[lb]{\cmd{if} %
            $\max\tset{d(u^{i,j},u^{i-1,j}), d(u^{i,j},u^{i+1,j}), d(u^{i,j},u^{i,j-1}), d(u^{i,j},u^{i,j+1})}<d_-$ \cmd{then}%
        }}
        \addtocounter{spc}{\vspc}
        \addtocounter{ind}{\istp} 
        \put(\theind,-\thespc){\makebox(0,0)[lb]{active$[u'']\leftarrow 0$}}
        \addtocounter{spc}{\vspc}
        \addtocounter{ind}{-\istp} 
        \put(\theind,-\thespc){\makebox(0,0)[lb]{\cmd{fi}}}
        \addtocounter{spc}{\vspc}
        \addtocounter{ind}{-\istp} 
        \put(\theind,-\thespc){\makebox(0,0)[lb]{\cmd{od}}}
        \addtocounter{spc}{\vspc}
        \put(\theind,-\thespc){\makebox(0,0)[lb]{t $\leftarrow$ t+1}}
        \addtocounter{spc}{\vspc}
        \addtocounter{ind}{-\istp} 
        \put(\theind,-\thespc){\makebox(0,0)[lb]{\cmd{od}}}
      }
    \end{picture}
  \end{center}
  \caption{\capstyle The training algorithm with limited dynamic adaptation of
    the architecture. The thick bars indicate a portion of the code
    inside a loop that is executed in parallel.}
  \label{trainingalg}
\end{figure}

The training algorithm, shown in Figure~\ref{trainingalg}, works on
the data of Table~\ref{training_input}. For training, each unit $u$
has an attribute called active$[u]$ which determines whether the unit
is active or not. These attributes are initialized to form one of the
topologies if Figure~\ref{initialization} (such initialization is not
shown).
\newcounter{spc}
\newcounter{ind}
When a document $D$ arrives, it is represented as a point $p$ in the
word space using the techniques considered in the previous
section. The algorithm that decides whether the document should be
displayed in shown in Figure \ref{displayalg}, while
Table~\ref{displaytable} shows the quantities used by the
algorithm.
\section{Testing Method}
\subsection{Data set}
The method presented here is evaluated on a data set derived from the
\emph{Reuters Corpus Volume 1} (RCV1-v1) \cite{lewis:04}, a collection
of $806,791$ news stories in NewsML format \cite{newsml:12} created by
Reuters journalists over a period of a year. For each document,
category codes for topic, region and industry sector are identified
and assigned as corresponding meta data.

The RCV1-v1 test collection contains a total of $126$ topics,
distributed in a hierarchy with, at the top level, four general areas:
\textbf{CCAT} (Corporate/Industrial), \textbf{ECAT} (Economics),
\textbf{GCAT} (Government/Social), and \textbf{MCAT} (Markets). Each
general area is the root of a topic sub-tree of depth at most 2. Of
the $126$ topics that compose the hierarchy, $23$ had no news assigned
to them. These topics were removed from the hierarchy, and all the
results presented here refer to the remaining $103$ topics.
\begin{table}[htb]
  \begin{center}
    \begin{tabular}{|l|p{4in}|}
      \hline
      $\som = \tset{u^{ij} | 1 \le i,j \le N_s}$ & the trained network containing the context; \\
      \hline
      $D$ & the document that just arrived \\
      \hline
      $p$ & the representation of $D$ in the word space \\
      \hline
      $L=\tlist{(D_1,s_1),\ldots,(D_L,s_L)}$ & the documents currently displayed with the respective scores, ordered such that $s_i>s_{i+1}$; \\
      \hline
      $\rho(d)$ & the urgency relaxation function; \\
      \hline
    \end{tabular}
  \end{center}
  \caption{\capstyle Input elements for the result display algorithm.}
  \label{displaytable}
\end{table}
%
%
%
%
%
%
\begin{figure}[htbp]
  \begin{center}
    \setcounter{spc}{0}
    \setcounter{ind}{0}
    \def\istp{20}
    \def\vspc{13}
    \def\vspcc{20}
    \def\tsize{180}
    \addtocounter{spc}{\vspc}
    \setlength{\unitlength}{0.1em}
    \begin{picture}(470,\tsize)(0,-\tsize)
      \multiput(0,-\tsize)(470,0){2}{\line(0,1){\tsize}}
      \multiput(0,-\tsize)(0,\tsize){2}{\line(1,0){470}}
      \put(20,0){
        \put(\theind,-\thespc){%
          \makebox(0,0)[lb]{$u^*\leftarrow\argmax\tset{\lambda^{ij}s(p,u^{ij}) | u^{ij}\in\mbox{\som}}$}
        }
        \addtocounter{spc}{\vspc}
        \addtocounter{ind}{\istp}
        \put(\theind,-\thespc){%
          \makebox(0,0)[lb]{\cmd{foreach} $m^{i,j}$ \cmd{in} $M$ \cmd{do}}%
          \multiput(1,-2)(5,0){2}{\line(0,-1){80}}%
          \multiput(1,-2)(0,-80){2}{\line(1,0){5}}
        }
        \addtocounter{spc}{\vspc}
        \addtocounter{ind}{\istp}
        \put(\theind,-\thespc){%
          \makebox(0,0)[lb]{\cmd{if} $\delta(u^{ij},u^*)\le\Delta$ \cmd{then}}
        }
        \addtocounter{spc}{\vspcc}
        \addtocounter{ind}{\istp}
        \put(\theind,-\thespc){%
          \makebox(0,0)[lb]{$\displaystyle \displaystyle \lambda^{ij}\leftarrow\Bigl[1 + \bigl( \frac{1}{\Theta}-1 \bigr) \rho(\delta(u^{ij},u^*)) \Bigr] \lambda^{ij}(t)$}
        }
        \addtocounter{spc}{\vspc}
        \addtocounter{ind}{-\istp}
        \put(\theind,-\thespc){%
          \makebox(0,0)[lb]{\cmd{else}}
        }
        \addtocounter{spc}{\vspcc}
        \addtocounter{ind}{\istp}
        \put(\theind,-\thespc){%
          \makebox(0,0)[lb]{$\displaystyle\lambda^{i,j}\leftarrow\min\Bigl\{ 1, \lambda^{ij}(t) + \frac{\Theta-1}{K\Theta}\Bigr\}$}
        }
        \addtocounter{spc}{\vspc}
        \addtocounter{ind}{-\istp}
        \put(\theind,-\thespc){%
          \makebox(0,0)[lb]{\cmd{end}}
        }
        \addtocounter{spc}{\vspc}
        \addtocounter{ind}{-\istp}
        \put(\theind,-\thespc){%
          \makebox(0,0)[lb]{\cmd{od}}
        }
        \addtocounter{spc}{\vspc}
        \put(\theind,-\thespc){%
          \makebox(0,0)[lb]{\cmd{if} $s(p,u^*)>s_L$ \cmd{then}}
        }
        \addtocounter{spc}{\vspc}
        \addtocounter{ind}{\istp}
        \put(\theind,-\thespc){%
          \makebox(0,0)[lb]{$k\leftarrow$ index s.t. $s_k>s(p,u^*)\ge{s_{k+1}}$}
        }
        \addtocounter{spc}{\vspc}
        \put(\theind,-\thespc){%
          \makebox(0,0)[lb]{$L\leftarrow\tlist{(D_1,s_1),\ldots,(D_k,s_k),(D,s(p,u^*)),(D_{k+1},s_{k+1}),\ldots,(D_{L-1},s_{L-1})}$}
        }
        \addtocounter{spc}{\vspc}
        \addtocounter{ind}{-\istp}
        \put(\theind,-\thespc){%
          \makebox(0,0)[lb]{\cmd{fi}}
        }
      }
    \end{picture}
  \end{center}
  \caption{\capstyle The algorithm that filters and displays the incoming documents.}
  \label{displayalg}
\end{figure}
The experiments were carried out using three different contexts, that
is, three different maps trained on one of the sub-topics of the
general areas \textbf{MCAT}, \textbf{GCAT}, and \textbf{CCAT}. The
area \textbf{ECAT} was not used as it represents a very small portion
of the collection and has a strong overlap with \textbf{CCAT} and
\textbf{MCAT}, resulting in very noisy data, not suited for measuring
performance. The data about the general areas and the data that were
used to build the context are reported in Table~\ref{contextbuild}.
\begin{table}[h]
  \begin{center}
    \begin{tabular}{|c|c|l|c|c|}
      \cline{3-5}
      \multicolumn{2}{c}{\textbf{}}  & \multicolumn{3}{|c|}{\textbf{The Context}} \\ 
      \hline
      \textbf{General Area} & \rule[-1.5em]{0pt}{1.5em}\parbox{5em}{\textbf{\% of}\\\textbf{total}} & 
      \textbf{Topics} & \rule[-1.5em]{0pt}{1.5em}\parbox{5em}{\textbf{\% in}\\\textbf{context}} & \textbf{news \#} \\
      \hline
      \textbf{MCAT} & 22  & M11 (EQUITY MARKETS)     & 5  & 5100 \\
                          \cline{3-4}
                          &     & M13 (MONEY MARKETS)     & 5   &                        \\ 
      \hline
      \textbf{GCAT} & 25  & GDIP (International relations)     & 5  & 4415 \\
                          \cline{3-4}
                          &     & GPOL (Domestic Politics)     & 5   &                        \\ 
      \hline
      \textbf{CCAT} & 41  & C17 (Funding/Capital)        & 5  & 4100 \\
                          \cline{3-4}
                          &     & C31 (Markets/Marketing)     & 5   &                        \\ 
      \hline
    \end{tabular}
  \end{center}
  \caption{\capstyle The top categories of the Reuters collection and their use
    for the creation of context. For each category, we show the
    percentage of the total corpus represented by that category. For
    each category, we have chosen two sub-topics and from each one we
    have drawn 5\% of the news as a training set for the network (the
    three training set have been used to train three separate
    networks). For instance, {\bf MCAT} represents a total of about
    $22\%$ of the news stories corpus. To create the context {\bf
      MCAT} we used $5\%$ of the news documents of each of its
    sub-topics {\bf M11} and {\bf M13}, for a total of about $5100$
    documents.}
  \label{contextbuild}
\end{table}
For each area, we selected two sub-topics, and 5\% of the news in each
sub-topic was randomly selected to train the context. The choice of
\emph{two} sub-topics represents a compromise: on the one hand,
drawing news from a variety of sub-topics creates a more general
context, one that better represents the general characteristics of the
area; on the other hand, using too many will not give us a good idea
of the capacity of the method to detect news related to, but not too
similar to, the news that are part of the context. The decision of
using two sub-topics was validated during preliminary tests.

The three general areas are not independent of each other: some
articles belong to several topics in more than one area. Since we
shall use the general areas to measure precision (an item will be a
\emph{hit} if it comes from the same area as the context), this
overlap imposes a limit on the performance that we can hope to
obtain. Figure~\ref{confusion} shows the confusion matrices between
the three general areas.
\begin{figure}[bhtp]
  \begin{center}
    \setlength{\unitlength}{1em}
    \begin{picture}(30,12)(0,0)
      \put(0,0){
        \multiput(1,0)(0,3){4}{\line(1,0){11}}
        \multiput(3,0)(3,0){4}{\line(0,1){11}}
        \put(2.5,1.5){\makebox(0,0)[r]{\textbf{GCAT}}}
        \put(2.5,4.5){\makebox(0,0)[r]{\textbf{CCAT}}}
        \put(2.5,7.5){\makebox(0,0)[r]{\textbf{MCAT}}}
        \put(4.5,9.5){\makebox(0,0)[b]{\textbf{MCAT}}}
        \put(7.5,9.5){\makebox(0,0)[b]{\textbf{CCAT}}}
        \put(10.5,9.5){\makebox(0,0)[b]{\textbf{GCAT}}}
        \put(4.5,1.5){\makebox(0,0){34.9}}
        \put(7.5,1.5){\makebox(0,0){41.1}}
        \put(10.5,1.5){\makebox(0,0){100}}
        \put(4.5,4.5){\makebox(0,0){47.7}}
        \put(7.5,4.5){\makebox(0,0){100}}
        \put(10.5,4.5){\makebox(0,0){51.6}}
        \put(4.5,7.5){\makebox(0,0){100}}
        \put(7.5,7.5){\makebox(0,0){57.6}}
        \put(10.5,7.5){\makebox(0,0){56.5}}
        \put(6,-0.5){\makebox(0,0){(a) word confusion.}}
        }
      \put(18,0){
        \multiput(1,0)(0,3){4}{\line(1,0){11}}
        \multiput(3,0)(3,0){4}{\line(0,1){11}}
        \put(2.5,1.5){\makebox(0,0)[r]{\textbf{GCAT}}}
        \put(2.5,4.5){\makebox(0,0)[r]{\textbf{CCAT}}}
        \put(2.5,7.5){\makebox(0,0)[r]{\textbf{MCAT}}}
        \put(4.5,9.5){\makebox(0,0)[b]{\textbf{MCAT}}}
        \put(7.5,9.5){\makebox(0,0)[b]{\textbf{CCAT}}}
        \put(10.5,9.5){\makebox(0,0)[b]{\textbf{GCAT}}}
        \put(4.5,1.5){\makebox(0,0){1}}
        \put(7.5,1.5){\makebox(0,0){7}}
        \put(10.5,1.5){\makebox(0,0){100}}
        \put(4.5,4.5){\makebox(0,0){11}}
        \put(7.5,4.5){\makebox(0,0){100}}
        \put(10.5,4.5){\makebox(0,0){7}}
        \put(4.5,7.5){\makebox(0,0){100}}
        \put(7.5,7.5){\makebox(0,0){4}}
        \put(10.5,7.5){\makebox(0,0){3}}
        \put(6,-0.5){\makebox(0,0){(b) news confusion.}}
        }
    \end{picture}
  \end{center}
  \caption{\capstyle Confusion matrices for the three general areas that are
    used in our experiments; (a): percentage of stems that appear in
    more than one area (e.g. 57.6\% of the word stems that appear in
    \textbf{MCAT} also appear in \textbf{CCAT}); (b): percentage of
    the news items that appear in more than one area (e.g. 4\% of the
    items that are classified as belonging to \textbf{MCAT} are
    classified as belonging to \textbf{CCAT} as well).}
  \label{confusion}
\end{figure}
Figure~\ref{confusion}.(a) shows the percentage of stems that appear
in more than one category (e.g. 57.6\% of the word stems that appear
in \textbf{MCAT} also appear in \textbf{CCAT});
Figure~\ref{confusion}.(b) shows the percentage of the news items that
appear in more than one area (e.g. 4\% of the items that are
classified as belonging to \textbf{MCAT} are classified as belonging
to \textbf{CCAT} as well). Note that while the overlapping in terms of
items is relatively small, reaching 10\% of an area only in one case
(\textbf{CCAT} vs.\ \textbf{MCAT}), the overlapping in words is
considerable, being almost always greater than 40\%.

This entails that, on the one hand, the judged similarity between
these areas is quite small: most articles are judged by the Reuters
journalist to belong to one and only one area but, on the other hand,
the statistical properties of the language used are much more
confusing: all these areas, especially \textbf{CCAT} and \textbf{MCAT}
use a similar vocabulary so that classification based on word
distribution only is extremely difficult.

\subsection{Training}
For each of the three general areas, a context was created using a
similar procedure. The new items taken at random from two sub-topics
are processed using the general method outlined in Section
\ref{sec:somActiveApproach}. As mentioned in the comments to
(\ref{sentence}), the unit that we use to create points in the
word space is the sentence: each sentence in the document is mapped to
a point in the input space. This decision was the result of some
preliminary tests that used bi-grams \cite{mayfield:98}, sentences and
paragraphs as units.

Each of these contexts (\textbf{MCAT}, \textbf{CCAT}, and
\textbf{GCAT}) has been used in nine different tests with nine
different network models resulting from three options for updating
the interest factor (constant interest with $\lambda^{ij}\equiv{1}$,
\emph{drastic}, and \emph{graded} with $\Delta=2$) and the three
options for the network architecture (fixed architecture, interwoven,
and checkerboard).  In the figures, The resulting possibilities will
be indicated with the codes of Table~\ref{codes}.
\begin{table}
  \begin{center}
    \begin{tabular}{p{10em}|c|c|c|}
      \multicolumn{1}{r|}{\makebox(0,0)[br]{architecture}} & 
      \multicolumn{3}{|c}{urgency} \\
      \cline{2-4}
        &  SOM &  Drastic & Graded \\
        & $\lambda\equiv{1}$ &  & $\Delta_0=2$ \\
      \cline{2-4}
      Fixed &  \nsym{SF} & \nsym{DF} & \nsym{GF} \\
      \cline{2-4}
      Adaptive &           &             &             \\
      \multicolumn{1}{r|}{checkerboard}  &  \nsym{SAC} & \nsym{DAC} & \nsym{GAC} \\
      \multicolumn{1}{r|}{alternate}     &  \nsym{SAA} & \nsym{DAA} & \nsym{GAA} \\
      \cline{2-4}
    \end{tabular}
  \end{center}
  \caption{\capstyle Codes used in the following for the various combinations of
    urgency factor and network architecture that we shall use in the
    tests. These symbols will be prefixed with the area used to train
    the context, so \nsym{GCAT/SF} refers to a standard (fixed)
    architecture without interest update ($\lambda\equiv{1}$) trained
    on \nsym{GCAT}, \nsym{MCAT/GAC} refers to a checkerboard dynamic
    architecture with graded interest update ($\Delta=2$) trained on
    \nsym{MCAT}, and so on.}
  \label{codes}
\end{table}
In addition to these, we shall use the codes \textbf{MCAT},
\textbf{CCAT}, and \textbf{GCAT} to indicate the category that had
been used to build the context. So the results relative to the network
for the MCAT context, with fixed architecture and drastic interest
reduction will be indicated as \textbf{MCAT/DF}, and so on.

The active network uses two parameters, the \emph{maximum activation
  distance} $d_+$, and the minimum deactivation distance $d_-$, which
depend on the \emph{context diameter} $d_\phi$ (see
eq. (\ref{eqMaxActivationDist}--\ref{eqMinDeactivationDist})). The
value $d_\phi$ was measured for the three contexts that we were used,
and the values $d_+$ and $d_-$ computed accordingly. The results are
shown in Table~\ref{activations}a.
\begin{table}
  \begin{center}
    \begin{tabular}{ccc}
      \begin{tabular}{|r|c|c|c|}
        \hline
        Context & $d_\phi$ & $d_+$ & $d_-$ \\
        \hline
        \hline
        \textbf{MCAT} & 0.59 & 0.29 & 0.05 \\
        \hline
        \textbf{CCAT} & 0.61 & 0.30 & 0.06 \\
        \hline
        \textbf{GCAT} & 0.67 & 0.33 & 0.06 \\
        \hline
      \end{tabular}
      & & 
      \begin{tabular}{|r|c|}
        \hline
        $\phi_m$ & 10  \\
        \hline
        $\sigma$ & 5 \\
        \hline
        $N_s$  & 30 \\
        \hline
        MAXITER & 10,000 \\
        \hline
      \end{tabular} \\
      (a) & \rule{2em}{0pt} & (b) 
    \end{tabular}
  \end{center}
  \caption{\capstyle (a): The values of the context diameter $d_\phi$ for the
    three contexts used in the tests and the resulting values of the
    maximum activation distance $d_+$ and the minimum deactivation
    distance $d_-$; these values correspond to setting $\mu=2$ and
    $\nu=10$ in (\ref{eqInitWeightsActivation}) and
    (\ref{eqMinDeactivationDist}).  (b): The learning parameters used
    in the tests. With these values, the initial value of the training
    parameter $\zeta$ is about 0.9, and the distance from the BMU at
    which learning practically doesn't take place is 5.}
  \label{activations} 
\end{table}

They correspond to selecting $\mu=2$ and $\nu=10$ in
(\ref{eqInitWeightsActivation}) and
(\ref{eqMinDeactivationDist}). The other learning parameters are shown
in Table~\ref{activations}b.

\subsection{Measurements}
\label{measect}
We base all our measurement on a \emph{testing list} consisting, at
any time, of the 500 best scoring items. The size of the list is kept
fixed (when a new item arrives which is to be inserted into the list,
the last item of the list is eliminated). At the beginning of
each run of the system there is a transient during which the list
fills up; all the measurements that we report are averages obtained
once the system reaches a steady state.

The basic non-novelty performance evaluation that we use is the
\emph{precision}, defined as the fraction of elements in a list that
belong to the same topic as the one used for the creation of the
context. We don't measure recall as the fixed size of the testing
list, and its relatively small size with respect to the size of each
category prevents the system from obtaining a significant recall.

A second element that we consider is the \emph{coverage}. Coverage is
a form of diversity measure: it measures how much of the context has
been covered by a group of news. It is defined as 
\begin{equation}
  \mbox{cov}(n) = \frac{\mbox{NBMU}(n)}{n}
\end{equation}
where $\mbox{NBMU}(n)$ is the number of different units that have
been selected as BMUs by the last $n$ items that have
arrived. We always compute this value with a number $n$ smaller than the
number of units in the map, so that $\mbox{cov}=1$ is theoretically
possible.

\section{Results}
\begin{figure}
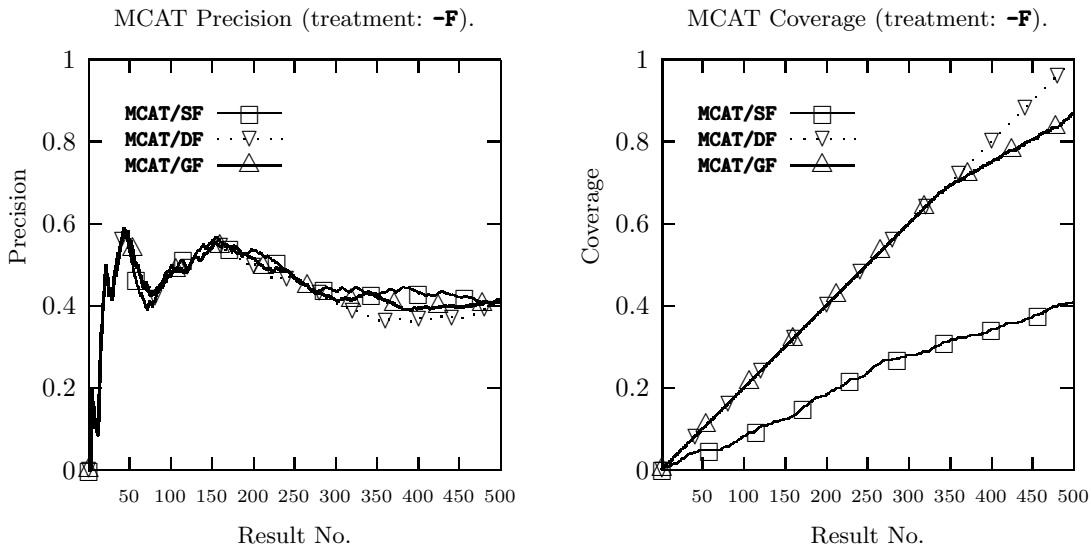

  \begin{center}
\setlength{\unitlength}{0.240900pt}
\ifx\plotpoint\undefined\newsavebox{\plotpoint}\fi
\sbox{\plotpoint}{\rule[-0.200pt]{0.400pt}{0.400pt}}%

  \end{center}
  \caption{\capstyle Precision and coverage for \nsym{MCAT/SF},
    \nsym{MCAT/DF}, \nsym{MCAT/GF}}
  \label{MCAT_F}
\end{figure}

\begin{figure}
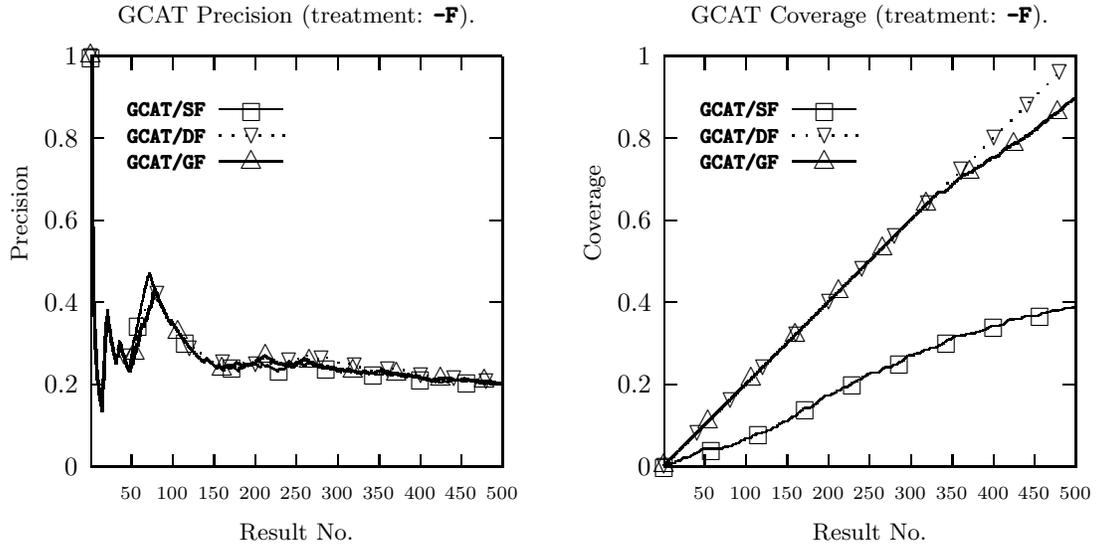

  \begin{center}
\setlength{\unitlength}{0.240900pt}
\ifx\plotpoint\undefined\newsavebox{\plotpoint}\fi
%
  \end{center}
  \caption{\capstyle Precision and coverage for \nsym{GCAT/SF},
    \nsym{GCAT/DF}, \nsym{GCAT/GF}}
  \label{GCAT_F}
\end{figure}

\begin{figure}
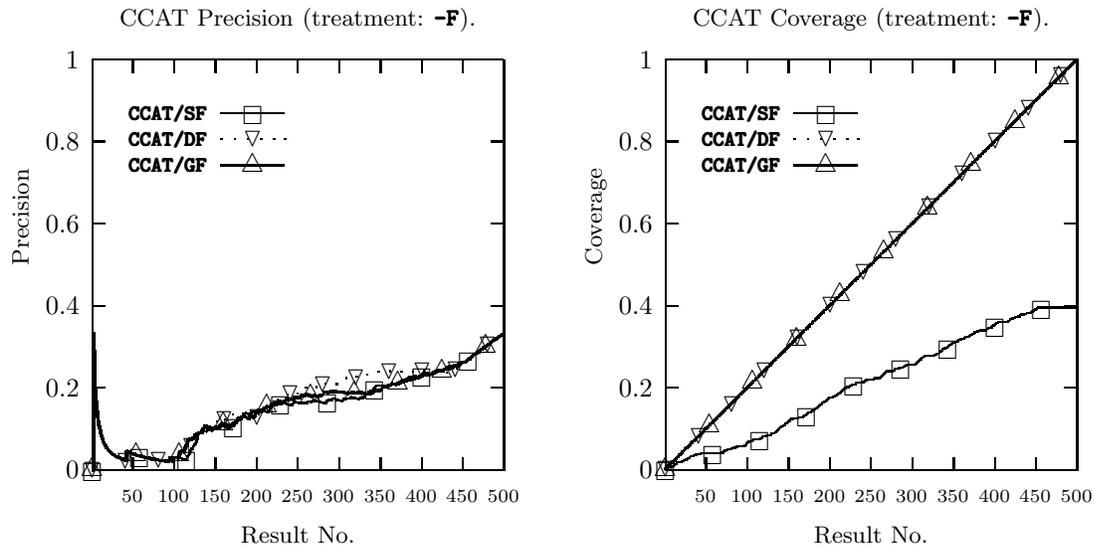

  \begin{center}
  \end{center}
\setlength{\unitlength}{0.240900pt}
\ifx\plotpoint\undefined\newsavebox{\plotpoint}\fi
%
  \caption{\capstyle Precision and coverage for \nsym{CCAT/SF},
    \nsym{CCAT/DF}, \nsym{CCAT/GF}}
  \label{CCAT_F}
\end{figure}

All our tests are conducted using the Reuters Corpus Volume 1
collection \cite{terrier:06}. Based on these data, we measure the
precision achieved in the nine cases of Table~\ref{codes} together
with their coverage. The two are reported side-by-side because, in
general, an increase in coverage comes at the expense in precision, as
it has normally observed in the novelty literature. What we have to
verify in order to assess the quality of a method is that the increase
in coverage doesn't come at the expense of a large drop in
precision. That is, we expect that the tests labeled \nsym{--/D-} and
\nsym{--/G-} (those that use the interest factor) will have
considerably higher coverage than the tests labeled \nsym{--/S-} and
that they will have the same or just slightly lower
precision. Figures~\ref{MCAT_F} to \ref{CCAT_F} show the mean
precision and the mean coverage score at a depth of up to 500 items in
the result list for contexts with fixed architecture under the three
modes of urgency updates.
Figures~\ref{MCAT_C} to \ref{CCAT_C} show the same measures as
obtained when the context is implemented as an adaptive network,
initialized using the checkerboard scheme on the left of
Figure~\ref{initialization}.
\begin{figure}
  \begin{center}
\setlength{\unitlength}{0.240900pt}
\ifx\plotpoint\undefined\newsavebox{\plotpoint}\fi
\sbox{\plotpoint}{\rule[-0.200pt]{0.400pt}{0.400pt}}%
%
  \end{center}
  \caption{\capstyle Precision and coverage for \nsym{MCAT/SAC},
    \nsym{MCAT/DAC}, \nsym{MCAT/GAC}; the active network used the
  \emph{checkerboard} initialization scheme (left of
  Figure~\ref{initialization}).}
  \label{MCAT_C}
\end{figure}
\begin{figure}
  \begin{center}
\setlength{\unitlength}{0.240900pt}
\ifx\plotpoint\undefined\newsavebox{\plotpoint}\fi
%
  \end{center}
  \caption{\capstyle Precision and coverage for \nsym{MCAT/SAC},
    \nsym{MCAT/DAC}, \nsym{MCAT/GAC}; the active network used the
  \emph{checkerboard} initialization scheme (left of
  Figure~\ref{initialization}).}
  \label{GCAT_C}
\end{figure}
\begin{figure}
  \begin{center}
\setlength{\unitlength}{0.240900pt}
\ifx\plotpoint\undefined\newsavebox{\plotpoint}\fi
%
  \end{center}
  \caption{\capstyle Precision and coverage for \nsym{CCAT/SAC},
    \nsym{CCAT/DAC}, \nsym{CCAT/GAC}; the active network used the
  \emph{checkerboard} initialization scheme (left of
  Figure~\ref{initialization}).}
  \label{CCAT_C}
\end{figure}
Finally, Figures~\ref{MCAT_A} to \ref{CCAT_A} show the same measures
as obtained when the context is implemented as an adaptive network and
the adaptive network was initialized using the alternating lines
scheme on the right of Figure~\ref{initialization}.

\begin{figure}
  \begin{center}
\setlength{\unitlength}{0.240900pt}
\ifx\plotpoint\undefined\newsavebox{\plotpoint}\fi
\sbox{\plotpoint}{\rule[-0.200pt]{0.400pt}{0.400pt}}%

  \end{center}
  \caption{\capstyle Precision and coverage for \nsym{MCAT/SAA},
    \nsym{MCAT/DAA}, \nsym{MCAT/GAA}; the active network used the
  \emph{alternating lines} initialization scheme (right of
  Figure~\ref{initialization}).}
  \label{MCAT_A}
\end{figure}
\begin{figure}
  \begin{center}
\setlength{\unitlength}{0.240900pt}
\ifx\plotpoint\undefined\newsavebox{\plotpoint}\fi
%
  \end{center}
  \caption{\capstyle Precision and coverage for \nsym{GCAT/SAA},
    \nsym{GCAT/DAA}, \nsym{GCAT/GAA}; the active network used the
  \emph{alternating lines} initialization scheme (right of
  Figure~\ref{initialization}).}
  \label{GCAT_A}
\end{figure}
\begin{figure}
  \begin{center}
\setlength{\unitlength}{0.240900pt}
\ifx\plotpoint\undefined\newsavebox{\plotpoint}\fi
%
  \end{center}
  \caption{\capstyle Precision and coverage for \nsym{CCAT/SAA},
    \nsym{CCAT/DAA}, \nsym{CCAT/GAA}; the active network used the
  \emph{alternating lines} initialization scheme (right of
  Figure~\ref{initialization}).}
  \label{CCAT_A}
\end{figure}

The general trend that emerges from all the results is that the
increase in coverage comes virtually at no expense in precision. All
the graphs on the left of Figures~\ref{MCAT_F}--\ref{CCAT_A} show no
statistically significant difference between the \nsym{--/S-} (WEBSOM
without interest factor) and the \nsym{--/D-} or \nsym{--/G-}
treatments. The non-significance of the differences is confirmed by
ANOVA ($p<0.01$). In the MCAT and GCAT contexts, precision follows a
similar pattern: after a transition characterized by a significant
instability, precision reaches a peak and then begins a slow descent.

The origin of this behavior is, admittedly, not entirely clear. It is
possible that the characteristics of the data set have something to do
with the reduction in precision in large result lists. Such decline
may be due to the fact that, as we move down the list, we are
displaying items progressively farther away from the map, creating
more opportunities for elements from other categories to be taken in.

The \nsym{--/D-} and \nsym{--/G-} treatments do have, as expected,
quite an impact on coverage. All the \nsym{--/DF} and \nsym{--/GF}
treatments show a steady increase in coverage, reaching, for the
\nsym{--/DF} treatment, values close to one. The \nsym{--/SF}
treatments, without the urgency factor, shows significantly lower
coverage. This behavior is consistent for the three contexts,
\nsym{MCAT}, \nsym{GCAT}, \nsym{CCAT}. 

The adaptive treatments, \nsym{--/-AC} and \nsym{--/-AA}, show a
coverage significantly lower than the fixed architecture (treatments
\nsym{--/-F}). Our experiments did not allow us to give a definite
explanation of this phenomenon, but some judicious conjecture is
possible. The adaptive networks are dense in the areas from which most
of the items of the training set are drawn. Given that the collection
is quite homogeneous, this is also the area that most items of the
test set will activate. Some of these units will have been
deactivated by a low coverage but, at the same time, some of the
units in the area will have recovered so we enter in a dynamic in
which items belonging to this area will often find at least some
active unit that they will be close to. That is, the areas with high
density will be active most of the time, reducing the number of times
in which less dense areas are activated and therefore reducing
coverage.

From these results---partial as they may be---it appears that forms of
adaptive architecture could be used to compromise between a high
coverage and a high representation of topics of interest.

\section{Related work}
The techniques used in this paper represent a development with many
roots in different areas of computing. The basic document
representation is a variation from standard techniques in information
retrieval, from stemming to the vector space representation
\cite{salton:88}. We should note, however, that our technique can be
applied to any representation of documents in a vector space. In
particular, representations such as Word2Vec \cite{lilleberg:15} are
ideal candidates for our method.

The context representation is quite clearly related to WEBSOM
\cite{kaski:97}, which is itself an application of Kohonen's
self-organizing map \cite{kohonen:90a,kohonen:90}. Our interpretation
of it as a latent semantic manifold, which we have developed elsewhere
\cite{santini:08b,santini:09a} is based on an analogy with the
well-known technique of \emph{latent semantic analysis}
\cite{deerwester:90}, which can be seen as a linear version of our
non-linear map. The analogy is justified by the probabilistic
interpretation of the map carried out, for the limit case in which the
units form a continuum, in \cite{santini:96a}.

The concepts of novelty and diversity entered the information
retrieval literature towards the end of the 1990s.  There is a general
consensus that the ball started rolling in 1998, with a two-page
position paper by Carbonell and Goldstein \cite{carbonell:98}.

Diversity and novelty are often treated together, but they address
different concerns: diversity is offered as a solution for query
\emph{ambiguity}, due to the inherent ambiguity of language (a query
like \dqt{Manhattan} may refer to a borough of New York, a movie, an
indian tribe, or a drink), while novelty addresses query
\emph{underspecification} (if the query is, say, about the movie,
there are several aspects of it that a person may be interested in)
\cite{clarke:09}.

Xu and Yin \cite{xu:08} operate a quadripartite division of possible
systems along two axes. The first axis is presentation, and the
systems are divided as having \emph{compansatory} or \emph{step}
presentations. In a compensatory system, diversity and novelty are
considered together in order to provide a composite relevance
score. In a step system, relevance is considered first, as a gauge:
only documents that score above a certain threshold are
retained. Diversity is considered next, and is used to reorder the set
of documents that has passed the first gauge. The second axis deals
with interaction, and distinguishes between \emph{undirected} and
\emph{directed} systems. Undirected systems are \dqt{one-shot}: they
receive a query and return a list of result, returning at each
position a document that minimizes the redundancy with those already
returned. Directed systems receive an input from the user, indicating
in which area(s) she wants the search to continue. The paper presented
here doesn't fit exactly in any of these categories, as the problem
that we are solving is not within the parameters of Xu and Yin. In a
general sense, however, it can be considered as a compensatory,
directed system, in which direction comes from the user model.

A great variety of diversification methods have been developed for
information retrieval and recommendation systems. Some of them are
based on unstructured relevance, that is, they consider relevance as a
single numerical score that measures the fitness of a document as a
whole. Among these, \cite{wang:09} uses a model based on the financial
theory of portfolio diversification developed in \cite{markowitz:52},
while \cite{rafiei:10} specializes in web pages. Several methods
consider a document as expressing a number of topics, and diversity is
sought by increasing teh number of relevant topics expressed in the
result set; \cite{agrawal:09} is one of the best known examples of
this class of systems, but work along similar lines can be found in
\cite{radlinski:06,santos:10,welch:11}.

In recommendation systems, the focus shifts to more user-centered
systems, in which diversity is obtained by presenting items that cover
the various interests of the user. Topic models (also called, in this
milieu, \emph{aspect models}) are often used for user and item models
\cite{alfaya:15}. The techniques used in this case vary from Latent
Semantic Analysis \cite{zhai:03} to Latent Dirichlet Analysis (LDA)
\cite{carterette:09}, Facet-Model LDA \cite{he:12} or matrix
factorization \cite{vargas:12}. 

All these methods are based on a probabilistic interpretation of
relevance: the relevance of a document, $r\in[0,1]$ is interpreted as
the \emph{probability} that a user will find the document relevant. It
has been argued that in some cases a \emph{fuzzy} interpretation might
be more correct \cite{dubois:01}, in which case $r$ would be
interpreted as the \emph{degree} to which a document is considered
relevant. Some algorithms for diversification based on this idea has
been studied in \cite{santini:11d}. In addition to text documents,
diversification has been applied to image retrieval
\cite{santini:18b}.

Users have several criteria in mind when they talk about quality of
results, among which the most prominent seem to be \emph{topicality},
\emph{novelty}, \emph{ease of understanding}, \emph{reliability}, and
\emph{scope} \cite{xu:06}, although topicality seems to be the most
relevant criterion \shortcite{mizzaro:97}. These findings form the
foundations of the methodological division operated by the
\cite{xu:08}, but if a document is off-topic, all other factors are
irrelevant for judgment \shortcite{wang:98}. This property justifies
the study of \emph{step} systems, in which documents are ranked only
if they are beyond a certain threshold of topicality. On the other
hand, computing practitioners don't like arbitrary thresholds,
especially when the sensitivity of the system with respect to their
value is not easily evaluated, a circumstance that makes it sensible
to evaluate compensatory system as more practical and robust
solution. This practicality comes with a price: undirected system use
less information about the user and the problems they involve are
computationally harder.

Diversification is an inherently hard problem. Unlike the work
presented in this paper, with its emphasis on \dqt{soft} computing,
many diversification methods define objective functions that they try
to maximize. Unfortunately, the resulting problem is virtually always
NP-complete \cite{santini:11c} and approximate solutions---often in
the form of heavily sub-optimal greedy algorithms--- have to be found.

\section{Conclusions}
We have presented a collection of algorithms for filtering incoming
streams of documents to make them more relevant and interesting. We
recognize that the success of a system of this kind requires an
equilibrium between two contrasting needs. On the one hand, we want to
give preference to the documents that the user will find the most
interesting. On the other hand, we don't want a few topic to
monopolize the offering, making the system of limited interest and
usefulness.

We propose a reinterpretation of the standard concept of
\dqt{novelty}, a concept that has proved very useful for diversifying
results in information retrieval and recommendation systems. In our
conceptual schema, adapted to the presence of a user model and to
data that arrive as a stream, a document is \emph{novel} if it serves
a user need (as represented by the user model) that hasn't been
adequately served recently. A set of recently displayed items has good
\emph{coverage} if it represents a good fraction of the general user
interests. Coverage replaces, in our conceptual framework, the
traditional concept of diversity.

We model the user interests using a modification of WEBSOM. Our
algorithm uses sentences as the fundamental unit of signification, a
solution that gives us better results than traditional word count or
of fixed length $n$-grams. We also introduced a limited form of dynamic
adaptation to compensate in part the concentration of units around the
topics of high interest that is typical of self-organizing maps. It
has been shown that self-organizing maps lay their units in the input
space in a way that approximates the probability distribution of the
training set \cite{santini:96a}.  The dynamic adaptation can be seen as
a smoothing of this approximation, and our tests have shown its
effects on the dynamic behavior of the system, especially on the
recovery of interest for the parts of the user model that have
received interesting documents in the past.

In the standard definition of information retrieval, diversity is
obtained at the expense of relevance and, therefore, of
precision. Maximal theoretical precision in the Robertsonian model
would be achieved by repeating the most relevant document as many times
as it is necessary to fill the whole result list, and diversity can
only be obtained by reducing this theoretical maximum. This is not the
case in our conceptual framework. Our concept of precision is relative
to the whole gamut of a person's interests, and our notion of coverage
entails that such gamut of interests is well represented in the
documents that the person sees. That is, unlike diversity, it is
possible to achieve high coverage from within the set of relevant
documents, viz.\ without reducing precision. Our measurements confirm
that our method increases coverage without any statistically
significant drop in precision.

\end{document}